\documentclass[journal,comsoc]{IEEEtran}

\usepackage{graphicx}
\usepackage{epstopdf}
\usepackage{float}
\usepackage{algorithm,algorithmic}
\usepackage{array}
\usepackage{amsmath}
\usepackage{amssymb}
\usepackage{mdwmath}
\usepackage{CJK}
\usepackage{mdwtab}
\usepackage{eqparbox}
\usepackage{fixltx2e}
\usepackage{cases}
\usepackage{bm}
\usepackage{multirow}
\usepackage{psfrag}
\usepackage[usenames]{color}
\usepackage{mathtools}

\usepackage{fixmath}
\usepackage{mathdots}
\usepackage{latexsym}

\usepackage{url}
\usepackage{cite}
\ifCLASSOPTIONcompsoc
\usepackage[caption=false,font=normalsize,labelfont=sf,textfont=sf]{subfig}
\else
\usepackage[caption=false,font=footnotesize]{subfig}
\fi

\newcommand{\diag}{\mathop{\mathrm{diag}}}
\newcommand{\tr}{\mathop{\mathrm{Tr}}}

\makeatletter
\renewcommand*{\@opargbegintheorem}[3]{\trivlist
      \item[\hskip \labelsep{\bfseries #1\ #2}] \textbf{(#3):}\ }
\makeatother

\begin{document}

\title{Signal Shaping for Non-Uniform Beamspace Modulated mmWave Hybrid MIMO Communications}
\author{Shuaishuai~Guo,~\IEEEmembership{Member, IEEE,}
        Haixia~Zhang,~\IEEEmembership{Senior Member, IEEE,}
Peng~Zhang,~\IEEEmembership{Member, IEEE,}\\ Shuping~Dang,~\IEEEmembership{Member, IEEE,} Chengcheng Xu,
and   Mohamed-Slim Alouini,~\IEEEmembership{Fellow, IEEE}
\thanks{The work of S. Guo, and P. Zhang were supported in part by Young Taishan Scholars, in part by the National Natural Science Foundation of China under Grant 61801266 and in part by the Shandong Natural Science Foundation under Grant ZR2018BF033. The work of C. Xu and H. Zhang were supported  by  National Natural Science Foundation of China under Grant No. 61860206005 and No. 61671278 (\emph{Corresponding author: Haixia Zhang}).}
\thanks{S. Guo, H. Zhang and C. Xu are with Shandong Provincial Key Laboratory of Wireless Communication Technologies and School of Control Science and Engineering, Shandong University, Jinan 250061, China (email: shuaishuai\textunderscore guo@sdu.edu.cn; haixia.zhang@sdu.edu.cn; chengchengxu@mail. sdu.edu.cn).}
\thanks{P. Zhang is with the School of Computer Engineering, Weifang University, Weifang 261061, China (e-mail: sduzhangp@163.com).}
\thanks{S. Dang and M.-S. Alouini are with the Computer, Electrical and
Mathematical Science and Engineering Division,  King Abdullah University of Science and Technology (KAUST), Thuwal 23955-6900, Saudi Arabia (email: shuping.dang@kaust.edu.sa; slim.alouini@kaust.edu.sa).}

}

\maketitle
\begin{abstract}
This paper investigates adaptive signal shaping methods for millimeter wave (mmWave) multiple-input multiple-output (MIMO) communications based on the maximizing the minimum Euclidean distance (MMED) criterion. In this work, we utilize the indices of analog precoders to carry information and optimize the symbol vector sets used for each analog precoder activation state.  Specifically, we firstly propose a joint optimization based signal shaping (JOSS) approach,  in which the symbol vector sets used for all analog precoder activation states are jointly optimized by solving a series of quadratically constrained quadratic programming (QCQP) problems. JOSS exhibits good performance, however, with a high computational complexity. To reduce the computational complexity, we then propose a full precoding based signal shaping (FPSS) method and a diagonal precoding based signal shaping (DPSS) method, where the full or diagonal digital precoders for all analog precoder activation states are optimized by solving two small-scale QCQP problems. Simulation results show that the proposed signal shaping methods can provide considerable performance gain in reliability in comparison with existing mmWave transmission solutions.

\end{abstract}


\IEEEpeerreviewmaketitle

\section{Introduction}

\IEEEPARstart{M}{illimeter} wave (mmWave) communications are next frontier for wireless communications. As the signal frequency goes higher, the required antenna size becomes  smaller and a large number of antennas can be integrated in a limited area. Owing to the cost and hardware complexity, it is impractical to equip each antenna with a radio frequency (RF) chain.  As a result, multiple-input multiple-output (MIMO) systems with reduced  RF chains are becoming a new trend for mmWave MIMO communications, where hybrid precoding dividing the signal processing in analog and digital domains to reduce the number of RF chains has attracted a lot of attention.
Given such an mmWave hybrid MIMO system offering a fixed transmission rate of $n$  bits per channel use (bpcu),  we are interested in finding the optimal transmit vector set $\mathcal{X}_N=\{\mathbf{x}_1,\mathbf{x}_2,\cdots, \mathbf{x}_{N}\}$ ($N=2^n$) subject to an average power constraint to maximize the minimum Euclidean distances (MMED) among the noise-free received signal vectors. In this work, we also discuss the extension of the proposed signal shaping methods to other optimization criteria, including the minimizing the symbol error rate (MSER) criterion and the maximizing the mutual information (MMI) criterion.

\subsection{Related Work}

Transmit vectors of mmWave hybrid MIMO systems are jointly determined by the hybrid precoders and the symbol vectors. All existing precoding and symbol vector optimization approaches can be regarded as the signal shaping methods. To summarize, we classify related works into three categories according to the precoding strategy.

\subsubsection{Best Beamspace Based Signal Shaping (BBSS)} In this category,  only a couple of analog and digital precoders are employed at the transmitter to steer the beam to the best beamspace during the transmission in a coherent time slot. The signal shaping can be optimized by carefully designing the hybrid precoders. To do so, \cite{Ayach2014} proposed an orthogonal pursuit matching (OMP) based precoding in fully-connected hybrid (FCH) mmWave MIMO systems leveraging the channel sparsity.  To improve the spectral efficiency (SE), the authors of  \cite{Yu2016} developed alternating minimization algorithms for the hybrid precoder optimization.  
 \cite{Dai2015} and \cite{Han2015} investigated successive interference cancellation (SIC) based precoding in partially-connected hybrid (PCH) mmWave MIMO systems. By trading off the SE and implementation complexity, a hybrid precoder design with dynamic partially-connected MIMO structure was proposed in \cite{Park2017}. Recently, \cite{Huang2019} has developed a deep-learning-enabled mmWave massive MIMO framework for effective hybrid precoder optimization, where the hybrid precoders are selected through a training based deep neural network with a substantially reduced complexity. Considering existing hybrid precoding solutions typically require a large number of high-resolution phase shifters, which still suffer from high hardware complexity and power consumption. To address this issue, the authors of \cite{Li2019} employed a limited number of low-resolution phase shifters and an antenna switch network to realize the hybrid precoders.
It is worth mentioning that the hybrid precoder solutions to maximize the SE are based on a Gaussian input assumption, resulting in that the designs are far from the optimality in practical mmWave MIMO communications with finite alphabet inputs \cite{Wu2018}.
With practical finite alphabet inputs, \cite{Rajashekar2016,Wu2017,Jin2017} have recently developed various effective and efficient hybrid precoding methods to maximize the mutual information, which are referred as MMI precoding. However, it should be emphasized that the information carrying capability by changing the precoder activation state has not been explored by the BBSS approach, which promises the potential for further optimization.

\subsubsection{Uniform Beamspace Modulation Based Signal Shaping (UBMSS)} In this category, the information carrying capability by changing the precoder activation state has been explored by uniformly activating a set of precoders. For example, a receive spatial modulation (RSM) for line-of-sight (LOS) mmWave MIMO communication systems was proposed in \cite{Perovic2017}, where a set of precoders that steer the beams to each receive antenna were adopted. Later, a virtual space modulation (VSM) transmission scheme and hybrid precoder designs  were proposed in \cite{Lee2017, Wang2018}.  Using the sparse scattering nature of mmWave channel, \cite{Ding2017} proposed a spatial scattering modulation (SSM). Relying on the beam index for modulation, the authors of \cite{Ding2018} developed a beam index modulation (BIM) scheme and showed its superiority in SE for mmWave communications. Roughly speaking, the difference among above transmission schemes lies in that the employed analog and digital precoders are slightly different. They are the same in activating each beamspace with equal probabilities since the symbol vector sets used for all beamspace activation states are the same. This results in limited performance because different beamspaces corresponding to different channels have inherently different information carrying capabilities. Besides, the employed symbol vector set has not been optimized.

\subsubsection{Non-Uniform Beamspace Modulation Based Signal Shaping (NUBMSS)}  Most recently, we proposed a generalized non-uniform beamspace modulation (NUBM) for mmWave communications in \cite{Guo2019}, where the beamspace is activated more flexibly. In the proposed NUBM scheme,  good beamspaces are activated with  high probabilities while poor beamspaces are activated with low probabilities. It has been theoretically proven that NUBM proposed in  \cite{Guo2019} outperforms the best beamspace selection (BBS) approach in terms of SE.  It has also been proven in \cite{Guo2019a} that NUBM is capacity-achieving for MIMO communications subject to a limited number of RF chains. The analysis on SE is based on the Gaussian input assumption. With finite alphabet inputs in practice, the beamspaces can be activated with non-equal probabilities by employing different symbol vector sets for different analog precoder activation states, such as the adaptive modulation schemes studied in \cite{Gao2018,Gao2019}. However, the adaptive modulation schemes can only be chosen from a limited number of modulation orders, which are the power of two. How to optimize the input to each beamspace in the complex domain remains unsolved. In this paper, we target this problem. 

\subsection{Contributions}
The paper attempts to optimize the multi-dimensional symbol vector set for each beamspace activation state in the complex domain.  It is an intricate task since the multiple symbol vector set optimization couples the discrete set size optimization  and the continuous set entry optimization in the complex domain.

\begin{itemize}
\item Firstly,  we propose a joint optimization based signal shaping (JOSS) method, where the symbol vector sets used for each analog precoder activation state are optimized. The size of the sets are optimized in a recursive way. Given an optimized set size, the optimization of the entries in the sets is formulated as a quadratically constrained quadratic programming (QCQP) problem and can be solved by existing algorithms. JOSS exhibits good performance in reliability, however, with a high computational complexity.
\item Secondly, to reduce the complexity of JOSS, we then propose a full-precoding based signal shaping (FPSS) method and a diagonal-precoding based signal shaping (DPSS) method. Based on all adaptive modulation candidates, we refine the modulation symbol vector sets with full digital precoders or diagonal digital precoders. In our design, the full/diagonal digital precoders for each analog precoder activation state are different and jointly optimized by solving a small-scale QCQP problem.

\item Thirdly, comprehensive comparisons among JOSS, FPSS, and DPSS are made in terms of reliability and computational complexity. To show the superiority of the proposed designs over existing mmWave transmission solutions, we also compare the proposed signal shaping aided NUBM with BBSS and UBMSS in terms of minimum Euclidean distance and symbol error rate (SER).

\item Fourthly, we probe into the capability of the proposed signal shaping methods for mmWave hybrid MIMO systems in approaching the fully-digital signal shaping (FDSS) methods for mmWave fully-digital MIMO systems. Moreover, we investigate the impact of channel state information (CSI) estimation errors and hardware impairments on the performance by simulations. 
We discuss the extension to orthogonal frequency division multiplexing (OFDM) based mmWave broadband MIMO systems. In addition, the impact of hybrid receiver and the discussion on the implementation challenges are also included. 
\end{itemize}

\subsection{Organization and Notations}
The remainder of the paper is organized as follows. Section II describes the system model. Section III formulate the optimization problems. The proposed signal shaping methods are presented in Section IV. Section V  discusses the implementation challenges and  the extension to other criteria. Section VI presents the simulation results. Section VII concludes the paper.

In this paper, scalars are represented by italic lower-case letters. Boldface upper-case and lower-case letters are used to denote matrices and column vectors. 
$(\cdot)^T$ and $(\cdot)^H$ stand for the transpose and transpose-conjugate operations, respectively. 
$\operatorname{Tr}(\mathbf{A})$ and $\mathrm{rank}(\textbf{A})$ denote the trace and rank of matrix $\textbf{A}$, receptively. Furthermore, $||\textbf{A}||_{F}$ is the Frobenius norm of matrix $\textbf{A}$ and $\operatorname{diag}(\mathbf{A})$ denotes a vector formed by the diagonal elements of matrix $\textbf{A}$. 
For a vector $\textbf{a}$, $||\textbf{a}||_2$ denotes its $l_2$ norm. Moreover, $\operatorname{diag}(\mathbf{a})$ denotes a diagonal matrix whose diagonal elements are assigned by vector $\textbf{a}$. 
$\odot$ and $\otimes$ denote the Hadamard and Kronecker products. 
$\textbf{I}_N$ indicates the $N\times N$ identity matrix. $\textbf{0}_{n}$ and $\textbf{1}_n$ are $n$-dimensional all-zeros and all-ones vectors, repectively.
$\mathcal{CN}(\boldsymbol{\mu},\boldsymbol{\Sigma})$ denotes a complex Gaussian vector with mean $\boldsymbol{\mu}$ and covariance $\boldsymbol{\Sigma}$.
$\mathbb{C}$ denotes the set of complex numbers.
$\jmath$ represents the imaginary unit.
$\left \lfloor{\cdot}\right \rfloor$ denotes the floor operation. $\left(n\atop m\right)$ is a binomial coefficient. 
$\mathcal{U}^{M\times N} $ denotes the set of all $M\times N$-dimensional matrices whose elements have unit magnitude. For a set $\mathcal{A}$, $|\mathcal{A}|$ represents its size. 
$\log_2(\cdot)$ stands for the logarithmic functions of base $2$. For clarity, we tabulate all abbreviations in Table I and important notations in Table II.

\begin{table}[t] 
\centering
\caption{Summary of Abbreviations} 
\begin{tabular}{ | c || c |}
\hline
\hline
\textbf{Abbreviation} &  \textbf{Full name} \\ 
\hline
\hline
ADC & analog-to-digital converter  \\
\hline
AMSS & adaptive modulation-based signal shaping  \\
\hline
AoAs & angles of arrival \\
\hline
AoDs & angles of departure \\
\hline
AP & analog precoder\\
\hline
AWGN & additive white Gaussian noise  \\
\hline
BBSS & best beamspace based signal shaping  \\
\hline
BBS & best beamspace selection  \\
\hline
BIM & beam index modulation  \\
\hline
bpcu& bits per channel use\\
\hline
CSI & channel state information  \\
\hline
DAC& digital-to-analog converter\\
\hline
DPSS& diagonal-precoding based signal shaping\\
\hline
DP& digital precoder\\
\hline
EE & energy efficiency  \\
\hline
FCH & fully connected hybrid  \\
\hline
FDSS& fully-digital signal shaping\\
\hline
FPSS& full-precoding based signal shaping\\
\hline
GBM& generalized beamspace modulation\\
\hline
JOSS& joint optimization based signal shaping\\
\hline
LOS & Line-of-sight\\
\hline 
MIMO & multiple-input multiple-output \\
\hline
ML & maximum-likelihood\\
\hline 
MMED & maximizing the minimum Euclidean distance \\
\hline 
MMI & maximizing the mutual information \\
\hline 
mmWave & millimeter wave \\
\hline 
MSER & minimizing the symbol error rate\\
\hline 
MRC & maximum ratio combining\\
\hline
NUBM & non-uniform beamspace modulation  \\
\hline
NUBMSS & NUBM based signal shaping  \\
\hline
OFDM&  orthogonal frequency division multiplexing\\
\hline
OMP&orthogonal pursuit matching\\
\hline
PCH & partially connected hybrid  \\
\hline
QCQP & quadratically constrained quadratic programming  \\
\hline
RF & radio frequency  \\
\hline
RSM & receive spatial modulation \\
\hline
SE & spectral efficiency \\
\hline
SER & symbol error rate  \\
\hline
SIC&successive interference cancellation\\
\hline
SNR & signal-to-noise ratio  \\
\hline
SSM & spatial scattering modulation  \\
\hline
UBMSS & uniform beamspace modulation based signal shaping  \\
\hline
UPA & uniform planar array\\
\hline
VSM & virtual spatial modulation\\
\hline
\end{tabular}
\end{table}

\begin{table}[h] 
\centering
\caption{Summary of System Model Notations} 
\begin{tabular}{ | c || c |}
\hline
\hline
\textbf{Notation} &  \textbf{System Parameter} \\ 
\hline
\hline
$\mathbf{f}_t$, $\mathbf{f}_r$ & transmit and receive antenna array response vectors\\
\hline
$\{\textbf{F}_{RF}^k\}$ & analog precoders\\
\hline
$\{(\textbf{F}_{BB})^k_l\}$& digital precoders\\
\hline
$\mathbf{H}$& channel matrix\\
\hline
$K$& number of candidate analog precoders\\
\hline
${m}$& rank of the channel\\
\hline
${n}$& transmission rate in bpcu\\
\hline
$\mathbf{n}$& noise vector\\
\hline
$N$& number of transmit vectors\\
\hline
$N_t$& number of transmit antennas\\
\hline
$N_r$& number of receive antennas\\
\hline
$N_{RF}$& number of transmit radio frequency chains\\
\hline
$N_{RF}$& number of receive radio frequency chains\\
\hline
$P_\textbf{s}$& average power constraint for the symbol vectors\\
\hline
$P_\textbf{x}$& average power constraint for the transmit vectors\\
\hline
$\{\textbf{s}_l^k\}$&  symbol vectors\\
\hline
$\{\hat{\textbf{s}}_l^k\}$& precoded symbol vectors\\
\hline
$\{\mathbf{x}_i\}$ & transmit vectors \\
\hline
$\mathcal{X}_N$ &  transmit vector set\\
\hline
$\mathbf{y}$ & receive vector \\
\hline
$\{\mathcal{S}_k\}$ & symbol vector sets \\
\hline
$\mathcal{Z}$ & set of symbol vector sets \\
\hline
\end{tabular}
\end{table}

\section{System Model}
We consider a point-to-point mmWave MIMO system, where the transmitter has $N_{t}$ antennas and the receiver has $N_{r}$ antennas.
Let ${\mathbf{x}_i} \in \mathbb{C}^{N_{t}}$ denote the transmitted signal vector. Then, the received signal vector $\mathbf{y} \in \mathbb{C}^{N_{r}}$ can be presented as
\begin{equation}
\mathbf{y}=\mathbf{H} {\mathbf{x}_i}+\mathbf{n},
\end{equation}
where $\mathbf{n} \in \mathbb{C}^{N_{r}}$ denotes the additive white Gaussian noise (AWGN) vector with mean zero and variance $\sigma^{2}$ at the receiver, i.e., $\mathbf{n} \sim \mathcal{C} \mathcal{N}\left(\mathbf{0}_{N_{r}}, \sigma^{2} \mathbf{I}_{N_{r}}\right)$; $\mathbf{H} \in \mathbb{C}^{N_{r} \times N_{t}}$ is the channel matrix between the transmitter and the receiver. Due to the limited number of {scatterers} in the mmWave {propagation} environment, the commonly used rich-scattering model at low frequencies is no longer applicable. Here, we adopt the Saleh-Valenzuela model \cite{Saleh1987}, which is given by
\begin{equation}
\mathbf{H}=\frac{1}{\sqrt{L}} \sum_{l=1}^{L} \alpha_{l} \mathbf{f}_{r}\left(\theta_{l}^{r}, \phi_{l}^{r}\right) \mathbf{f}_{t}^{{H}}\left(\theta_{l}^{t}, \phi_{l}^{t}\right),
\end{equation}
where $L$ represents the number of effective propagation paths, and $\alpha_{l}$ is the channel coefficient of the $l$-th path. $\theta_{l}^{r} \in[0, \pi)$ and $\phi_{l}^{r} \in[0,2 \pi)$ are the elevation and azimuth angles of arrival (AoAs). $\theta_{l}^{t} \in[0, \pi]$ and $\phi_{l}^{t} \in[0,2 \pi]$ represent the elevation and azimuth angles of departure (AoDs). Finally, $\mathbf{f}_{t}\left(\theta_{l}^{t}, \phi_{l}^{t}\right)$ and $\mathbf{f}_{r}\left(\theta_{l}^{r}, \phi_{l}^{r}\right)$ denote the transmitter and receiver antenna array response vectors.
In this paper, an uniform planar array (UPA) with $W_{1}$ and $W_{2}$ elements ($W_{1}=W_{2}=\sqrt{N_w}$) on horizon and vertical is considered, whose array response vector can be written as
\begin{equation}
\begin{aligned} \mathbf{f}_{w}(\theta,\phi)=& \frac{1}{\sqrt{N_w}}\left[1, \ldots, e^{j \frac{2 \pi}{\lambda} d(x \sin (\phi) \sin (\theta)+y \cos (\theta))}, \ldots\right.\\ &\left.e^{j \frac{2 \pi}{\lambda} d\left(\left(W_{1}-1\right) \sin (\phi) \sin (\theta)+\left(W_{2}-1\right) \cos (\theta)\right)}\right]^{T}, \end{aligned}
\end{equation}
where $\lambda$ and $d$ represent the signal wavelength and the antenna spacing, respectively. In addition, $w=t$ or $r$ in $N_w$ and $\mathbf{f}_{w}(\theta,\phi)$, $0 \leq x \leq\left(W_{1}-1\right)$ and $0 \leq y \leq\left(W_{2}-1\right)$, where $x$ and $y$ stand for the antenna indices in the two-dimensional plane.

In this paper, we assume $\textbf{H}$ is perfectly known by the transceivers. It is noted that although this is an ideal assumption, in practical applications, CSI at the receiver can be obtained by the downlink channel estimation. Specifically, in time division duplex (TDD) systems with uplink and downlink channel reciprocity, CSI at the transmitter can be acquired by uplink channel estimation. In frequency division duplex (FDD) systems, CSI at the transmitter can be acquired by feeding back the estimated CSI from the receiver.

\begin{figure}[t]
  \centering
  \includegraphics[width=0.4\textwidth]{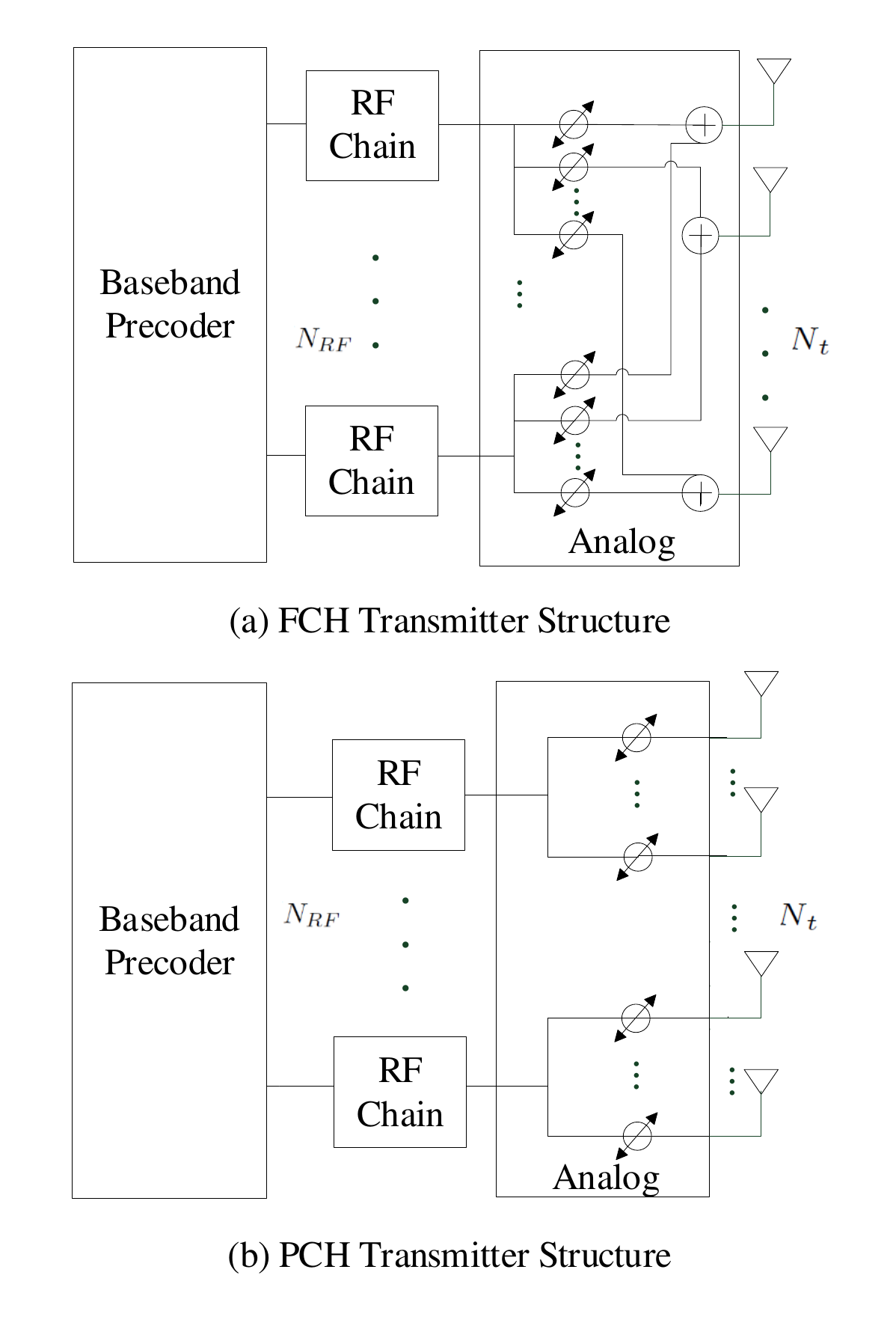}\\
 \caption{FCH and PCH mmWave MIMO transmitter structures.}
  \label{System_Model}
\end{figure}
In this work, we adopt the commonly considered hybrid analog and digital array architectures, which significantly reduce the number of required RF chains by cascading an analog feed network after the baseband digital signal processor. Fig. 1 depicts two major hybrid array architectures, namely the FCH array and PCH array. In both cases, the transmitter has $N_{t}$ antennas but only $N_{R F}$ ($\ll N_{t}$) RF chains and is capable of transmitting up to $N_{R F}$ independent data streams simultaneously\cite{Gao2016a,Jamali2019,Yan2019}.
In an FCH array architecture, each of $N_{R F}$ RF chains is connected to all $N_{t}$ antennas via $N_{t}$ phase shifters and an ($N_{R F}+1$)-port combiner. As a consequence, the fully-connected architecture provides full beamforming gain of massive antenna arrays but with a high hardware complexity of total $N_{R F} N_{t}$ phase shifters and $N_{t}$ combiners.
PCH architecture is also referred to as sub-array, where $N_{t}$ antennas are partitioned into $N_{t} / N_{R F}$ groups and each RF chain is connected to only a subset of antennas. Therefore, the number of required phase-shifters is reduced to $N_{t}$, and no power combiner is needed.
There exists a trade-off between energy efficiency (EE) and SE  for the two hybrid architectures. That is, the FCH architecture can provide the full beamforming gain at the expense of hardware cost/power consumption, whereas the low complexity PCH architecture realizes a low beamforming gain\cite{Ahmed2018}. Moreover, it is noteworthy that FCH and PCH arrays are chosen as examples to introduce our work and the proposed designs can be directly extended to mmWave communications with other array structures.

\section{Problem Formulation}
In this paper, we are interested in the optimization of the transmit vector set. That is, for an mmWave hybrid MIMO system with a target transmission rate of $n$ bpcu, we aim to find the optimal vector set
$\mathcal{X}_{N}=\{\textbf{x}_1,\textbf{x}_2,\cdots,\textbf{x}_{N}\}$, where $N=2^n$. Maximizing the minimum Euclidean distance among the noise-free received signal vectors is our target, where the minimum Euclidean distance can be expressed as
\begin{equation}
d_{\min}(\mathcal{X}_N,\textbf{H})=\min_{i,i'\in\mathcal{I},i\neq i'}||\textbf{H}\textbf{x}_i-\textbf{H}\textbf{x}_{i'}||_2.
\end{equation}
With a hybrid structure, a transmit vector $\textbf{x}_i\in \mathcal{X}_{N}$  can be expressed as
\begin{equation}\label{Xi}
\textbf{x}_i=\textbf{F}_{RF}^k\hat{\textbf{s}}_l^k,
\end{equation}
where $\hat{\textbf{s}}_l^k=(\textbf{F}_{BB})_l^k\textbf{s}_l^k$ can be regarded as a precoded symbol vector by the digital precoder $(\textbf{F}_{BB})_l^k$; $\textbf{F}_{RF}^k$ denotes the $k$-th analog precoder and $\hat{\textbf{s}}_l^k$ denotes the $l$-th precoded symbol vector when $\textbf{F}_{RF}^k$ is activated. We use  $\mathcal{F}_{RF}=\{\textbf{F}_{RF}^1,\textbf{F}_{RF}^2,\cdots,\textbf{F}_{RF}^K\}$ of size $K$ to denote the set of all analog precoder candidates. Sets $\mathcal{S}_1,\mathcal{S}_2,\cdots,\mathcal{S}_{K}$ of sizes $n_1,n_2,\cdots,n_K$ are used to denote the precoded  symbol vector sets when $\textbf{F}_{RF}^1,\textbf{F}_{RF}^2,\cdots,\textbf{F}_{RF}^K$ are activated, respectively.

\textbf{Remark:} Based on the denotations, we can clearly see the differences among BBSS, UBMSS and NUBMSS.  In BBSS, only a fixed precoder $\textbf{F}_{RF}^1$ is adopted and other precoders will not be activated. In other words, we have the precoded symbol vector set sizes $n_1=N$ and $n_2=n_3=\cdots=n_K=0$. In UBMSS,  a subset of $\hat{K}=2^{\lfloor\log_2K \rfloor}$ precoders in $\mathcal{F}_{RF}$ are uniformly activated to send equal-size  $\mathcal{S}_1,\mathcal{S}_2,\cdots,\mathcal{S}_{\hat{K}}$. That is, $n_1=n_2=\cdots=n_{\hat{K}}=N/\hat{K}$ and $n_{\hat{K}+1}=n_{\hat{K}+2}=\cdots=n_K=0$. In NUBMSS, all  precoders are non-uniformly activated subject to a size constraint $\sum_{k=1}^Kn_{k}=N$. From this perspective,  it is found that BBSS and UBMSS can be regarded as the special realizations of NUBMSS and the globally optimized NUBMSS will inherently outperform BBSS and UBMSS.

For convenience, we further define $\mathcal{Z}\triangleq\left\{\mathcal{S}_1,\mathcal{S}_2,\cdots,\mathcal{S}_{K}\right\}$ and the minimum Euclidean distance among the noise-free received signal vectors can be rewritten as
\begin{equation}
\begin{split}
d_{\min}(\mathcal{X}_N,\textbf{H})&=d_{\min}(\mathcal{F}_{RF},\mathcal{Z},\textbf{H})\\&=\min_{{{\textbf{F}^{k}_{RF}\hat{\textbf{s}}_l^k\neq \textbf{F}^{k'}_{RF}\hat{\textbf{s}}_{l'}^{k'}\atop \textbf{F}^{k}_{RF},\textbf{F}^{k'}_{RF}\in\mathcal{F}}\atop\hat{\textbf{s}}_l^k\in\mathcal{S}_k,~\hat{\textbf{s}}_{l'}^{k'}\in\mathcal{S}_{k'} }\atop\mathcal{S}_k,\mathcal{S}_{k'}\in\mathcal{Z}}||\textbf{H}(\textbf{F}_{RF}^{k}\hat{\textbf{s}}_l^k-\textbf{F}^{k'}_{RF}\hat{\textbf{s}}_{l'}^{k'})||_2,
\end{split}
\end{equation}
The signal shaping becomes a problem finding $\mathcal{F}_{RF}$ and $\mathcal{Z}$ to maximize $d_{\min}(\mathcal{F}_{RF},\mathcal{Z},\textbf{H})$ subject to a size constraint that
\begin{equation}
\sum_{k=1}^{K}n_k=N,\\
\end{equation}
 and  an average power constraint $P_{\textbf{s}}$ that
\begin{equation}
P(\mathcal{Z})
=\frac{1}{N}\sum_{k=1}^{K}\sum_{l=1}^{n_k}(\hat{\textbf{s}}_l^k)^H{\hat{\textbf{s}}_l^k}\leq P_{\textbf{s}}.
\end{equation}
Thus, the signal shaping problem for UBMSS in mmWave hybrid MIMO communications can be formulated as
\begin{equation}
\begin{split}
(\textbf{P1}):~~~~~\mathrm{Given}: &~\textbf{H},N\\
\mathrm{Find}:&~\mathcal{F}_{RF}=\{\textbf{F}_{RF}^1,\textbf{F}_{RF}^2,\cdots,\textbf{F}_{RF}^K\},\\&~\mathcal{Z}=\left\{\mathcal{S}_1,\mathcal{S}_2,\cdots,\mathcal{S}_K\right\}\\
\mathrm{Maximize}:&~d_{\min}^2(\mathcal{F}_{RF},\mathcal{Z},\textbf{H})\\
\mathrm{Subject~to}:&~\sum_{k=1}^{K}n_k=N,\\
&~\textbf{F}_{RF}^k\in \mathcal{U}^{N_t\times N_{RF}},\\
&~P(\mathcal{Z})\leq P_\textbf{s}. \\
\end{split}
\end{equation}

The variables $\mathcal{F}_{RF}$ and $\mathcal{Z}$ are coupled. To solve the problem, we have to decouple them. In this paper, we propose to firstly optimize $\mathcal{F}_{RF}$ and then find the optimal $\mathcal{Z}$ based on the optimized $\mathcal{F}_{RF}$.
Since the channel considered in this paper is sparse, each precoder in $\mathcal{F}_{RF}$ should not steer the beam to the zero space of $\textbf{H}$.  To guarantee this, we provide a singular matrix approximation approach, which can be described as follows. First, we perform singular value decomposition (SVD) as
$\textbf{H}=\textbf{U}\boldsymbol{\Lambda}\textbf{V}^H$, where $\mathbf{U}\in \mathbb{C}^{N_r\times m}$ is the left-singular matrix, $\boldsymbol{\Lambda}\in\mathbb{C}^{m\times m}$ is the diagonal matrix with $m=\rm{rank}(\textbf{H})$ non-zero singular values as diagonal entries and $\mathbf{V}\in\mathbb{C}^{N_t\times m}$ is the right-singular matrix.  Since the symbol vectors should be transmitted through the subspace expanded by the $\mathbf{V}$ and for $N_{RF}$ data steams, there are $\left(m\atop N_{RF}\right)$ subspace matrices, i.e., $K=\left(m\atop N_{RF}\right)$ and we denote the subspace matrices as $\textbf{F}_1,\textbf{F}_2,\cdots,\textbf{F}_{K}$.  The subspace matrix are implemented by fully-digital structures. In our work, we adopt analog precoders and digital precoders to approximate $\left\{\textbf{F}_1,\textbf{F}_2,\cdots,\textbf{F}_K\right\}$. The approximation can be performed by solving
\begin{equation}
\begin{split}
(\textbf{P2}):\min_{\{\textbf{F}_{RF}^{k}\},\{\textbf{F}_{BB}^{k}\}}~~&||\textbf{F}_k-\textbf{F}_{RF}^{k}\textbf{F}_{BB}^{k}||_{F}^2\\
\mathrm{subject~to:}~~& \textbf{F}_{RF}^{k}\in \mathcal{U}^{N_t\times N_{RF}},~\textbf{F}_{BB}^k\in \mathbb{C}^{N_{RF}\times N_{RF}},\\
&|| \textbf{F}_{RF}^{k}\textbf{F}_{BB}^{k}||_F^2=|| \textbf{F}_{k}||_F^2.
\end{split}
\end{equation}
The problem can be solved by numerous existing algorithms, e.g., the OMP algorithm for  FCH MIMO systems in \cite{Ayach2014} or the SIC algoritm for PCH MIMO systems in \cite{Dai2015}. Besides, we note that only $\{\textbf{F}_{RF}^{k}\}$ is useful in our designs and $\{\textbf{F}_{BB}^k\}$ is considered in the next step of optimizing $\hat{\textbf{s}}^k_l$. It is noteworthy that this paper just provides a way for designing $\mathcal{F}_{RF}$ and the following signal shaping methods in the paper are suitable for any feasible $\mathcal{F}_{RF}$.

Given a feasible analog  precoder set $\mathcal{F}_{RF}$, the signal shaping problem is reduced to find the symbol vector sets for different analog precoder activation states, which can be given by
\begin{equation}
\begin{split}
(\textbf{P3}):~~~~~\mathrm{Given}: &~\textbf{H},N,\mathcal{F}_{RF}\\
\mathrm{Find}:&~\mathcal{Z}=\left\{\mathcal{S}_1,\mathcal{S}_2,\cdots,\mathcal{S}_K\right\}\\
\mathrm{Maximize}:&~d_{\min}^2(\mathcal{F}_{RF},\mathcal{Z},\textbf{H})\\
\mathrm{Subject~to}:&~\sum_{k=1}^{K}n_k=N,\\
&~P(\mathcal{Z})\leq P_s. \\
\end{split}
\end{equation}

\textbf{Remark:} The determination of $K=\left(m\atop N_{RF}\right)$ is owing to that the number of mutually-independent non-zero subspace matrices is $\left(m\atop N_{RF}\right)$.
The rationale for using $\left(m\atop N_{RF}\right)$ subspace matrices instead of restricting to the subspace matrix corresponding to the strongest $N_{RF}$ singular vectors is that the differences between subspace matrices are also employed to enlarge the mutual Euclidean distances among the noise-free received signal vectors.  We note that even though there exist $K$ legitimate beamspaces,  it does not mean that all $K$ beamspaces will be activated during the transmission. Whether a subspace will be used or not is determined by whether the associated symbol vector set is a non-empty set or not. If the associated symbol vector set of a beamspace matrix is empty, the corresponding beamspace will not be activated during the transmission phase.
In this subsection, we propose to solve the original problem (\textbf{P1}) by solving two separate subproblems (\textbf{P2}) and (\textbf{P3}). That is, we firstly find the non-zero beamspace set formed by feasible analog precoders, and then optimize the symbol vectors for each beampsace activation states. However, it is difficult to provide a rigorous proof for the equivalence in splitting (\textbf{P1}) into (\textbf{P2}) and (\textbf{P3}). To investigate the capability of the splitting in approaching the optimal performance, we compare the proposed signal shaping methods by solving (\textbf{P2}) and (\textbf{P3}) with the signal shaping by directly solving a relaxed problem of (\textbf{P1}), i.e., 
\begin{equation}
\begin{split}
(\textbf{RP1}):~~~~~\mathrm{Given}: &~\textbf{H},N\\
\mathrm{Find}:&~\mathcal{X}_N=\{\textbf{x}_{1},\textbf{x}_2,\cdots,\textbf{x}_{N}\}\\
\mathrm{Maximize}:&~d_{\min}^2(\mathcal{X}_N,\textbf{H})\\
\mathrm{Subject~to}:
&~P(\mathcal{X}_N)\leq P_{\textbf{x}},
\end{split}
\end{equation}
where the hybrid structure is relaxed and the average power constraint $ P_{\textbf{x}}$ on the transmit vector set is given by
\begin{equation}
P(\mathcal{X}_N)=\mathbb{E}(||\textbf{x}||^2)=\frac{1}{N}\sum_{i=1}^{N}{\textbf{x}_i}^T{\textbf{x}_i}\leq P_\textbf{x}.
\end{equation}
Problem (\textbf{RP1}) can be regarded the formulation of the signal shaping for mmWave fully-digital MIMO systems, whose solution can provide a performance bound for the solution of (\textbf{P1}). The detailed discussion on the solution of (\textbf{RP1}) and the comparison are included in Sections V-B and VI-A, respectively. Next, we focus our attention on solving (\textbf{P3}) since (\textbf{P2}) can be solved by existing algorithms.

\section{Signal Shaping Methods}
Problem (\textbf{P3}) is a set optimization problem. It includes the set size optimization, i.e., finding the optimal $n_1, n_2,\cdots,n_K$ that satisfy the size constraint 
$\sum_{k=1}^{K}n_k=N$. After that, one still needs to perform set entry optimization, i.e., finding the optimal set entries in $\mathcal{S}_1, \mathcal{S}_2,\cdots,\mathcal{S}_{K}$. To solve the problem, we propose three signal shaping approaches in this section.

\subsection{Joint Optimization Based Signal Shaping (JOSS)}
The set size optimization is a discrete optimization satisfying $\sum_{k=1}^{K}n_k=N$ and $n_k\geq 0,k=1,2,\cdots,n_k$. According to the analysis in \cite{Guo2017}, there are $\left(N+K-1\atop K-1\right)$ feasible solutions, which is a large number. For instance, given $N=64$, i.e., $n=6$ bpcu, $K=|\mathcal{F}_{RF}|=10$, there are around $\left(73\atop 9\right)=9.7\times10^{10}$ feasible set size solutions. For each set size solution, one also needs to perform set entry optimization. Thus, exhaustive search over all feasible set size solutions is prohibitive. To solve this problem for practical systems, we resort to a greedy recursive design method which was firstly introduced in  \cite{Guo2017}. To introduce the recursive design method, we first define a matrix $\textbf{G}_N\in\mathbb{C}^{N_t\times NN_{RF}}$ by
\begin{equation}
\begin{split}
\textbf{G}_{N}&\triangleq\\&\left[\overbrace{\textbf{F}_{RF}^1,\cdots,\textbf{F}_{RF}^1}^{n_1},\overbrace{\textbf{F}_{RF}^2,\cdots,\textbf{F}_{RF}^2}^{n_2},\cdots,\overbrace{{\textbf{F}_{RF}^{K},\cdots,\textbf{F}_{RF}^K}}^{n_{K}}\right],
\end{split}
\end{equation}
which corresponds to a feasible solution $(n_1,n_2,\cdots,n_K)$. With the definition of $\textbf{G}_N$, the recursive design can be described as follows. Given $\textbf{G}_{N-1}$, we can choose an $\textbf{F}_{RF}^k\in \mathcal{F}_{RF}$ to adjoin $\textbf{G}_{N-1}$ for generating $|\mathcal{F}|$ candidates of $\textbf{G}_{N}$. For each candidate of $\textbf{G}_{N}$, we perform the set size optimization and obtain the corresponding candidates of $\mathcal{X}_N$. Then, by comparing all of the  candidates of $\mathcal{X}_N$, we can obtain a suboptimal $\mathcal{X}_N$ from all of the candidates and the corresponding suboptimal $\textbf{G}_{N}$.  According to this principle, we use the optimal $\mathcal{X}_{2}$ and $\textbf{G}_{2}$, which can be obtained by exhaustive search, to find a suboptimal $\mathcal{X}_{3}$ and $\textbf{G}_{3}$, then $\mathcal{X}_{4}$ and $\textbf{G}_{4}$ and so on, until the size constraint is satisfied.

For denotation convenience, we use $\textbf{G}$ to represent $\textbf{G}_N$. The set entry optimization in the recursive design can be performed as follows. We define ${\textbf{S}_l}^k\triangleq\diag(\hat{\textbf{s}}_l^k)$ for all $k=1,2,\cdots,K$, $l=1,2,\cdots,n_k$ and express the transmit vector $\textbf{x}_i$ as
\begin{equation}
\textbf{x}_i=\textbf{G}\textbf{D}_{\textbf{z}}\textbf{o}_i,
\end{equation}
where 
$\textbf{D}_{\textbf{z}}\in\mathbb{C}^{NN_{RF}\times NN_{RF}}$ is a diagonal matrix  defined by
\begin{equation}
\textbf{D}_{\textbf{z}}\triangleq 
\begin{bmatrix}
    ~\textbf{S}_1^1 & \textbf{0} & \textbf{0}&\cdots&\textbf{0}&\textbf{0}&  \textbf{0}~ \\
    ~\textbf{0} & \ddots& \textbf{0}& \cdots&\cdot&\textbf{0}& \textbf{0}~ \\
     ~\textbf{0} & \textbf{0} & \textbf{S}_{n_1}^1&  \cdots&\textbf{0}&\cdot&\textbf{0}~\\
~\vdots& \vdots & \vdots& \ddots& \vdots&\vdots&\vdots~\\
    ~\textbf{0} & \cdot & \textbf{0} &\cdots &\textbf{S}_{1}^{K}& \textbf{0}&  \textbf{0}~\\
 ~\textbf{0} & \textbf{0} & \cdot &\cdots &\textbf{0}& \ddots&  \textbf{0}~\\
 ~\textbf{0} & \textbf{0} & \textbf{0} &\cdots &\textbf{0}& \textbf{0}&  \textbf{S}_{n_K}^{K}~
\end{bmatrix},
\end{equation}
and  $\textbf{o}_i\in\mathbb{R}^{NN_{RF}}$ as $\textbf{o}_i\triangleq\textbf{g}_i\otimes \textbf{1}_{N_{RF}}$,  where $\textbf{g}_i$ is the $i$th $N$-dimensional vector basis with all zeros except the $i$th entry being one. 
Based on these definitions, the square of the pairwise Euclidean distances can be expressed as
\begin{equation}\label{eqD}
\begin{split}
||\textbf{H}\textbf{x}_i-\textbf{H}\textbf{x}_{i'}||_2^2&=||\textbf{H}\textbf{G}\textbf{D}_{\textbf{z}}\textbf{o}_i-\textbf{H}\textbf{G}\textbf{D}_{\textbf{z}}\textbf{o}_{i'}||_2^2\\
&=(\textbf{o}_i-\textbf{o}_{i'})^H\textbf{D}_\textbf{z}^H\textbf{G}^H\textbf{H}^H\textbf{H}\textbf{G}\textbf{D}_\textbf{z}(\textbf{o}_i-\textbf{o}_{i'})\\
&=\tr\left(\textbf{D}_\textbf{z}^H\textbf{R}_{\textbf{HG}}\textbf{D}_\textbf{z}\Delta\textbf{O}_{i{i'}}\right),
\end{split}
\end{equation}
where $\textbf{R}_{\textbf{H}\textbf{G}}=\textbf{G}^H\textbf{H}^H\textbf{H}\textbf{G}$ and $\Delta\textbf{O}_{i{i'}}=(\textbf{o}_i-\textbf{o}_{i'})(\textbf{o}_i-\textbf{o}_{i'})^H$.
Given any two diagonal matrices $\textbf{D}_{\textbf{u}}=\diag(\textbf{u})$ and $\textbf{D}_{\textbf{v}}=\diag(\textbf{v})$, we have an equality 
 $\tr(\textbf{D}_{\textbf{u}}\textbf{U}\textbf{D}_{\textbf{v}}\textbf{V}^H)=\textbf{u}^H(\textbf{U}\odot\textbf{V})\textbf{v}$,
based on which (\ref{eqD}) can be re-expressed as
\begin{equation}\label{eq18D}
||\textbf{H}\textbf{x}_i-\textbf{H}\textbf{x}_{i'}||_2^2=\textbf{z}^H\textbf{Z}_{i{i'}}\textbf{z},
\end{equation}
where $\textbf{z}= \diag\{\textbf{D}_\textbf{z}\}\in\mathbb{C}^{NN_{RF}}$ and $\textbf{Z}_{i{i'}}=\textbf{R}_{\textbf{H}\textbf{G}}\odot\Delta\textbf{O}_{i{i'}}^H\in\mathbb{C}^{NN_{RF}\times NN_{RF}}$.
As a consequence, the average power constraint can be expressed as
\begin{equation}
P(\mathcal{Z})
=\frac{1}{N}\tr\left(\textbf{D}_\textbf{z}\textbf{D}_\textbf{z}^H\right)
=\frac{1}{N}\textbf{z}^H\textbf{z}\leq P_\textbf{s}.
\end{equation}
Based on the above reformulations, the set entry optimization  becomes
\begin{equation}
\begin{split}
(\textbf{P4}):~~\mathrm{Given}: &~\textbf{Z}_{ii'}, \forall i\neq i'\in\{1,2,\cdots,N\}\\
\mathrm{Find}:&~\textbf{z}\\
\mathrm{Maximize}:&~\min \textbf{z}^H\textbf{Z}_{ii'}\textbf{z}\\
\mathrm{Subject~to}:&~\textbf{z}^H\textbf{z}\leq NP_\textbf{s}.
\end{split}
\end{equation}
Because the minimum Euclidean distance monotonically increases with the increase of the average power, maximizing the minimum Euclidean distance in (\textbf{P4}) can also be reformulated to minimize the average power for a target minimum distance $d_T$, which can be expressed by
\begin{equation}
\begin{split}
(\textbf{P5}):~\mathrm{Given}: &~\textbf{Z}_{ii'}, \forall i\neq i'\in\{1,2,\cdots,N\}\\
\mathrm{Find}:&~\textbf{z}\\
\mathrm{Minimize}:&~\textbf{z}^T\textbf{z}\\
\mathrm{Subject~to}:&~\textbf{z}^T\textbf{Z}_{ii'}\textbf{z}\geq d_T^2,\forall i\neq i'\in\{1,2,\cdots,N\}.
\end{split}
\end{equation}
It is worth mentioning that problem (\textbf{P5}) is formulated without any power constraint, and hence  the optimized transmit vectors should be further scaled to satisfy the average power constraint.
Problem (\textbf{P5}) is an optimization problem in which both the objective function and the constraints are quadratic functions. That is,  (\textbf{P5}) is a typical quadratically constrained quadratic programming (QCQP) problem with $NN_{RF}$ complex variables  and $\left(N\atop 2\right)$ constraints, which can be solved by existing algorithms, e.g., the algorithm in \cite{Cheng2018}, with a complexity of $\mathcal{O}(N^4N_{RF}^2)$.  For clearly viewing the recursive design process, we list the JOSS in Algorithm 1. The algorithm in \cite{Cheng2018} for solving the non-convex QCQP problems is a kind of gradient descent algorithm. It starts from a randomly generated solution and is updated when the objective function is decreased.  Since the objective function is lower bounded by $0$, the algorithmic convergence can thereby be ensured.  The convergence rate is high and  has been investigated in \cite{Cheng2018}. 

\begin{algorithm}[t]
\caption{JOSS algorithm}
\label{alg:A}
\begin{algorithmic}
\STATE  \textbf{Input:} $\textbf{H},N,\mathcal{F}_{RF}$ 
\STATE  \textbf{Output:} $\mathcal{X}_N$
\STATE {Generate $|\mathcal{F}_{RF}|^2$ feasible candidates of $\textbf{G}_2$, and compute $\textbf{z}_{2}$ by solving (\textbf{P5}) using the algorithm in \cite{Cheng2018}. Compare all  the candidates of $\mathcal{X}_2$, which are generated by  $\textbf{G}_2$  and $\textbf{z}_{2}$, and obtain the optimal $\mathcal{X}_2$ and the corresponding  $\textbf{G}_2$.}
\STATE {Initialize $t=3$.}
\REPEAT
\STATE{ Generate $|\mathcal{F}_{RF}|$ feasible $\textbf{G}_t$ based on $\textbf{G}_{t-1}$. and compute $\textbf{z}_{t}$ by solving (\textbf{P5}) using the algorithm in \cite{Cheng2018}. Compare all  the candidates of $\mathcal{X}_t$, which are generated by  $\textbf{G}_t$  and $\textbf{z}_{t}$, and obtain the optimal $\mathcal{X}_t$ and the corresponding  $\textbf{G}_t$.}
\STATE{Update} $t\leftarrow t+1$.
\UNTIL{$t>N$.}
\STATE{Output the optimized $\mathcal{X}_N$.} 
\end{algorithmic}
\end{algorithm}

According to similar complexity analysis in \cite{Guo2017,Guo2019b}, the aggregated computational complexity of JOSS is 
\begin{equation}\label{Complexity}
\mathcal{C}_{\rm{JO}}= \mathcal{O}(N_{\mathrm{iter}}^{\text{JO}}KN^5N_{RF}^2),
\end{equation}
where $N_{\mathrm{iter}}^{\text{JO}}$ denotes the average iteration number that the algorithm in \cite{Cheng2018} takes to converge to solve (\textbf{P5}).

\subsection{Full Precoding Based Signal Shaping (FPSS)}
Observing the computational complexity in  (\ref{Complexity}), it is found that the complexity is at least the fifth power of the transmit vector set size $N$. In the case with a large $N$, we have to resort to their methods for solving this problem. In this subsection, we propose the full precoding based signal shaping approach. The idea is expatiated as follows. First, we express $\textbf{x}_i\in \mathcal{X}_{N}$  as
\begin{equation}\label{XiN22}
\textbf{x}_i=\textbf{F}_{RF}^k\textbf{F}_{BB}^k\textbf{s}_l^k.
\end{equation}
We denote $\mathcal{S}_1^c, \mathcal{S}_2^c, \cdots, \mathcal{S}_K^c$ as unprecoded symbol vector sets which are optimized in a given codebook similarly to that in adaptive modulation schemes and $\mathcal{Z}_c=\{\mathcal{S}_1^c,\mathcal{S}_2^c,\cdots,\mathcal{S}_K^c\}$. It is slightly different from the expression in (\ref{Xi}). The difference lies in that we refine $\{\textbf{s}_l^k\}$ with the same $\textbf{F}_{RF}^k$, i.e., $(\textbf{F}_{BB})_1^k=(\textbf{F}_{BB})_2^k=\cdots(\textbf{F}_{BB})_{n_c}^k=\textbf{F}_{RF}^k$, while in (\ref{Xi}) $\{\textbf{s}_l^k\}$  are respectively precoded by different $\{(\textbf{F}_{BB})_l^k\}$. The new expression can be regarded as a special case of the general case in (\ref{Xi}). 
The optimized performance is inherently less comparable to the globally optimized solution. But, the most important thing is that the expression can reduce the optimization complexity. The detailed optimization procedure is described as follows.

\begin{table*}[t]
\centering
\caption{Comparisons among different signal shaping methods}\label{tab1}
    \begin{tabular}{ | c |c| c |c|c|}
    \hline
Approach&Parameters to be optimized&Number of variables &Number of constraints&Computational complexity\\
\hline
JOSS in Section IV-A & $\{\hat{\textbf{s}}_l^k\}=\{(\textbf{F}_{BB})_l^k\textbf{s}_l^k\}$& $NN_{RF}$&$\left(N\atop2\right)$&$\mathcal{O}(N_{\mathrm{iter}}^{\text{JO}}KN^5N_{RF}^4)$\\
\hline
FPSS  in Section IV-B &Full $\{\textbf{F}_{BB}^k\}$& $KN_{RF}^2$&$\left(N\atop2\right)$&$\mathcal{O}\left(N_cN_{\mathrm{iter}}^{\mathrm{FP}}K^2N^2N_{RF}^4\right)$\\
\hline
DPSS  in Section IV-C& Diagonal $\{\hat{\textbf{F}}_{BB}^k\}$& $KN_{RF}$&$\left(N\atop2\right)$&$\mathcal{O}\left(N_cN_{\mathrm{iter}}^{\mathrm{DP}}K^2N^2N_{RF}^2\right)$\\
\hline
FDSS  in Section V-B & Fully-digital $\{\textbf{x}_i\}$&$NN_{t}$&$\left(N\atop2\right)$&$\mathcal{O}\left(N_{\mathrm{iter}}^{\mathrm{FD}}N^4N_{t}^2\right)$\\
\hline
    \end{tabular}
\end{table*}

Given $\mathcal{Z}_c$ which is chosen from feasible normalized modulation symbol vector sets for adaptive modulation schemes, we optimize the digital precoders $\textbf{F}_{BB}^1,\textbf{F}_{BB}^2,\cdots,\textbf{F}_{BB}^K$ to refine $\mathcal{S}_1^c,\mathcal{S}_2^c,\cdots,\mathcal{S}_K^c$ in $\mathcal{Z}_c$. By denoting $n_1^c,n_2^c,\cdots,n_K^c$ as the set sizes of $\mathcal{S}_1^c,\mathcal{S}_2^c,\cdots,\mathcal{S}_K^c$, we re-express $\textbf{x}_i$ in (\ref{XiN22}) as
\begin{equation}\label{NXi}
\textbf{x}_i=\textbf{F}_{RF}^k\textbf{F}_{BB}^k\textbf{s}_l^k=\textbf{W}\textbf{D}_{\textbf{q}}\textbf{e}_i,
\end{equation}
where the matrix $\textbf{W}$ of size $N_t\times KN_{RF}^2$ is defined as
\begin{equation}
\textbf{W}\triangleq\left[\overbrace{\textbf{F}_{RF}^1,\cdots,\textbf{F}_{RF}^1}^{N_{RF}},\overbrace{\textbf{F}_{RF}^2,\cdots,\textbf{F}_{RF}^2}^{N_{RF}},\cdots,\overbrace{{\textbf{F}_{RF}^{K},\cdots,\textbf{F}_{RF}^K}}^{N_{RF}}\right],
\end{equation}
 the diagonal matrix $\textbf{D}_{\textbf{q}}$ of size $KN_{RF}^2\times KN_{RF}^2$ is defined as 
\begin{equation}
\textbf{D}_{\textbf{q}}\triangleq 
\begin{bmatrix}
    ~\textbf{Q}_{BB}^1 & \textbf{0} &\cdots&\textbf{0}\\
    ~\textbf{0} & ~\textbf{Q}_{BB}^2& \cdots& \textbf{0}~ \\
~\vdots& \vdots &  \ddots& \vdots~\\
    ~\textbf{0} &0 &\cdots &\textbf{Q}^{K}_{BB}\\
\end{bmatrix},
\end{equation} 
the diagonal matrix $\textbf{Q}_i$ of size $N_{RF}^2\times N_{RF}^2$ is defined as
\begin{equation}
\begin{split}
&\textbf{Q}_{BB}^k\\&=\diag\left(\textrm{vec}\left(\left(\textbf{F}_{BB}^k\right)^T\right)\right)\\
&=\text{diag}\left\{\left[\left(\textbf{F}_{BB}^k\right)_{1,1}, \left(\textbf{F}_{BB}^k\right)_{1,2},\cdots,\left(\textbf{F}_{BB}^k\right)_{N_{RF}, N_{RF}}\right]^T\right\},
\end{split}
\end{equation}
the vector $\textbf{e}_i$ of size $KN_{RF}^2\times 1$ can be expressed as
\begin{equation}
\textbf{e}_i=\tilde{\textbf{g}}_k\otimes\tilde{\textbf{s}}_l^k,
\end{equation}
the vector $\tilde{\textbf{s}}_l^k$ of size $N_{RF}^2\times 1$ is expressed as
\begin{equation}
 \tilde{\textbf{s}}_l^k=[\overbrace{({\textbf{s}}_l^k)^T,({\textbf{s}}_l^k)^T,\cdots,({\textbf{s}}_l^k)^T}^{N_{RF}}]^T,
\end{equation}
and $\tilde{\textbf{g}}_k$ is a $K$-dimensional vector basis with all zeros except the $k$th entry being one. Because the proof of the reformulation is the same as that in \cite{Cheng2018}, we do not repeat it here for simplicity.  

With the reformulation in  (\ref{NXi}), we rewrite the square of the Euclidean distance between $\textbf{H}\textbf{x}_i$ and $\textbf{H}\textbf{x}_{i'}$ as
\begin{equation}\label{eq11}
\begin{split}
||\textbf{H}\textbf{x}_i-\textbf{H}\textbf{x}_{i'}||^2&=||\textbf{H}\textbf{W}\textbf{D}_\textbf{q}(\textbf{e}_i-\textbf{e}_{i'})||^2\\
&=(\textbf{e}_i-\textbf{e}_{i'})^H\textbf{D}_\textbf{q}^H\textbf{W}^H\textbf{H}^H\textbf{H}\textbf{W}\textbf{D}_\textbf{q}(\textbf{e}_i-\textbf{e}_{i'})\\
&=\tr\left(\textbf{D}_\textbf{q}^H\textbf{R}_{\textbf{HW}}\textbf{D}_\textbf{q}\Delta\textbf{E}_{i{i'}}\right),
\end{split}
\end{equation}
where $\textbf{R}_{\textbf{HW}}=\textbf{W}^H\textbf{H}^H\textbf{H}\textbf{W}$ and $\Delta\textbf{E}_{i{i'}}=(\textbf{e}_i-\textbf{e}_{i'})(\textbf{e}_i-\textbf{e}_{i'})^H$.
Similarly, we can rewrite (\ref{eq11}) as
\begin{equation}\label{XiN}
||\textbf{H}\textbf{x}_i-\textbf{H}\textbf{x}_{i'}||^2=\textbf{q}^T\textbf{Q}_{i{i'}}\textbf{q},
\end{equation}
where $\textbf{q}= \diag\{\textbf{D}_\textbf{q}\}\in\mathbb{C}^{KN_{RF}^2}$ and $\textbf{Q}_{i{i'}}=\textbf{R}_{\textbf{HW}}\odot\Delta\textbf{E}_{i{i'}}^T\in\mathbb{C}^{KN_{RF}^2\times KN_{RF}^2}$.
Accordingly, the power constraint can be given by
\begin{equation}\label{C1}
P(\mathcal{Z})
=\frac{1}{N}\tr\left(\textbf{D}_\textbf{q}\textbf{D}_\textbf{q}^H\right)
=\frac{1}{N}\textbf{q}^H\textbf{q}\leq 1.
\end{equation}
According to (\ref{XiN}) and (\ref{C1}), the original problem can be expressed as
\begin{equation}
\begin{split}
(\textbf{P6}):~~\mathrm{Given}: &~\textbf{Q}_{ii'}, \forall i\neq i'\in\{1,2,\cdots,n_k+1\}\\
\mathrm{Find}:&~\textbf{q}\\
\mathrm{Maximize}:&~\min \textbf{q}^T\textbf{Q}_{ii'}\textbf{q}\\
\mathrm{Subject~to}:&~\textbf{q}^T\textbf{q}\leq N.
\end{split}
\end{equation}
By introducing an auxiliary variable $\tau$, similarly to Section IV-A,  the optimization problem can be transformed to be
\begin{equation}
\begin{split}
(\textbf{P7}):~\mathrm{Given}: &~\textbf{Q}_{ii'}, \forall i\neq i'\in\{1,2,\cdots,N\}\\
\mathrm{Find}:&~\textbf{q}\\
\mathrm{Minimize}:&~\textbf{q}^T\textbf{q}\\
\mathrm{Subject~to}:&~\textbf{q}^T\textbf{Q}_{ii'}\textbf{q}\geq \tau,\forall i\neq i'\in\{1,2,\cdots,N\}.
\end{split}
\end{equation}
It is a QCQP problem with $KN_{RF}^2$ variables and $\left(N\atop 2\right)$ constraints. The problem can also be solved by using existing algorithms, e.g., the one in \cite{Cheng2018}. The corresponding computational complexity is around $\mathcal{O}(N_{\mathrm{iter}}^{\mathrm{FP}}K^2N^2N_{RF}^4)$, where $N_{\mathrm{iter}}^{\text{FP}}$ denotes the average iteration number that the algorithm in \cite{Cheng2018} takes to converge to solve (\textbf{P7}). For all candidates for $\mathcal{Z}_c$, we preform the refinement, compare them in terms of the minimum Euclidean distance, and then obtain the final signal shaping. The aggregated computational complexity can thereby be expressed as $\mathcal{O}\left(N_cN_{\mathrm{iter}}^{\mathrm{FP}}K^2N^2N_{RF}^4\right)$, where $N_c$ denotes the number of feasible candidates for $\mathcal{Z}_c$ in the adaptive modulation scheme.

\subsection{Diagonal Precoding Signal Shaping (DPSS)}
Besides employing the full digital precoders $\textbf{F}_{BB}^1,\textbf{F}_{BB}^2,\cdots,\textbf{F}_{BB}^K$ to refine $\mathcal{S}_1^c,~\mathcal{S}_2^c,\cdots,~\mathcal{S}_K^c$, one can also employ diagonal precoders $\hat{\textbf{F}}_{BB}^1,~\hat{\textbf{F}}_{BB}^2,~\cdots,~\hat{\textbf{F}}_{BB}^K$ to perform the refinement, which can reduce the optimization and implementation complexity. Similarly, the refinement can be performed as follows.
First, we define a matrix $\hat{\textbf{W}}$ of size $N_t\times KN_{RF}$ as
\begin{equation}
\hat{\textbf{W}}\triangleq\left[\textbf{F}_{RF}^1,\textbf{F}_{RF}^2,\cdots,\textbf{F}_{RF}^{K}\right],
\end{equation}
 a diagonal matrix $\hat{\textbf{D}}_{\textbf{q}}$ of size $KN_{RF}\times KN_{RF}$ as
\begin{equation}
\hat{\textbf{D}}_{\textbf{q}}\triangleq 
\begin{bmatrix}
    ~\hat{\textbf{F}}_{BB}^1 & \textbf{0} &\cdots&\textbf{0}\\
    ~\textbf{0} & ~\hat{\textbf{F}}_{BB}^2& \cdots& \textbf{0}~ \\
~\vdots& \vdots &  \ddots& \vdots~\\
    ~\textbf{0} &0 &\cdots &\hat{\textbf{F}}^{K}_{BB}\\
\end{bmatrix},
\end{equation} 
and a vector $\hat{\textbf{e}}_i$ of size $KN_{RF}\times 1$ as
\begin{equation}
\hat{\textbf{e}}_i=\tilde{\textbf{g}}_k\otimes\textbf{s}_l^k.
\end{equation}
Then, the transmit vector in use of diagonal precoders can be expressed by
\begin{equation}\label{DNXi}
\textbf{x}_i=\textbf{F}_{RF}^k\hat{\textbf{F}}_{BB}^k\textbf{s}_l^k=\hat{\textbf{W}}\hat{\textbf{D}}_{\textbf{q}}\hat{\textbf{e}}_i.
\end{equation}
Similarly, we can directly optimize $\hat{\textbf{F}}_{BB}^1,~\hat{\textbf{F}}_{BB}^2,~\cdots,~\hat{\textbf{F}}_{BB}^K$ jointly by solving
\begin{equation}
\begin{split}
(\textbf{P8}):~\mathrm{Given}: &~\hat{\textbf{Q}}_{ii'}, \forall i\neq i'\in\{1,2,\cdots,N\}\\
\mathrm{Find}:&~\hat{\textbf{q}}\\
\mathrm{Minimize}:&~\hat{\textbf{q}}^T\hat{\textbf{q}}\\
\mathrm{Subject~to}:&~\hat{\textbf{q}}^T\hat{\textbf{Q}}_{ii'}\hat{\textbf{q}}\geq \tau,\forall i\neq i'\in\{1,2,\cdots,N\},
\end{split}
\end{equation}
where $\hat{\textbf{q}}= \diag\{\hat{\textbf{D}_\textbf{q}}\}\in\mathbb{C}^{KN_{RF}}$, $\hat{\textbf{Q}}_{i{i'}}=\hat{\textbf{R}}_{\textbf{H}\hat{\textbf{W}}}\odot\Delta\hat{\textbf{E}}_{i{i'}}^T\in\mathbb{C}^{KN_{RF}\times KN_{RF}}$, $\hat{\textbf{R}}_{\textbf{H}\hat{\textbf{W}}}=\hat{\textbf{W}}^H\textbf{H}^H\textbf{H}\hat{\textbf{W}}$ and $\Delta\hat{\textbf{E}}_{i{i'}}=(\hat{\textbf{e}}_i-\hat{\textbf{e}}_{i'})(\hat{\textbf{e}}_i-\hat{\textbf{e}}_{i'})^H$. 

It is also a QCQP problem with $KN_{RF}$ variables and $\left(N\atop 2\right)$ constraints. The corresponding computational complexity is around $\mathcal{O}(N_{\mathrm{iter}}^{\mathrm{DP}}K^2N_{RF}^2N^2)$, where $N_{\mathrm{iter}}^{\text{FP}}$ denotes the average iteration number, by which the algorithm in \cite{Cheng2018} takes to converge to solve (\textbf{P8}). For all candidates for $\mathcal{Z}_c$, the aggregated computational complexity for refinement can thereby be expressed as $\mathcal{O}\left(N_cN_{\mathrm{iter}}^{\mathrm{DP}}K^2N^2N_{RF}^2\right)$.

\subsection{Comparison}
 The proposed signal shaping approaches are quite similar since the optimizations are all conducted by solving QPCP problems and the number of constraints are the same. But, the numbers of their optimization variables are  different, which induce difference in computational complexity. 
For clearly viewing their differences, we illustrate them in Table III. 


\section{Extension and Discussion}
The most unique characteristic of mmWave communications is the broadband nature. To benefit from the broadband nature, we discuss the signal shaping for OFDM-based mmWave MIMO communications. Besides, we also discuss the performance loss compared to fully-digital signal shaping (FDSS), implementation challenges, the design with a hybrid receiver structure, the extension to MSER and MMI signal shaping methods in this section.

\subsection{Extension to Broadband mmWave Communications}
Using OFDM, the proposed signal shaping methods can be directly extended to broadband mmWave MIMO systems. Particularly, let $\tilde{k}$ represent the sub-carrier index and $\tilde{K}$ be the number of carriers, the received signal in the frequency domain can be given by
\begin{equation}
\textbf{y}[\tilde{k}]=\rho\textbf{H}[\tilde{k}]\textbf{F}_{RF}^k\hat{\textbf{s}}_l^k[\tilde{k}]+\textbf{n}[\tilde{k}], \tilde{k}=0,2,\cdots,\tilde{K}-1,
\end{equation}
where $\textbf{H}[k]$ denotes the channel matrix of the $\tilde{k}$-th sub-carrier, which is also characterized  by the Saleh-Valenzuela model \cite{Yu2016}
\begin{equation}\label{HK}
\begin{split}
\textbf{H}[\tilde{k}]=\frac{1}{\sqrt{L}}\sum_{l=1}^{L} \alpha_{l}
\mathbf{f}_{r}\left(\theta_{l}^{r}, \phi_{l}^{r}\right) \mathbf{f}_{t}^{\mathrm{H}}\left(\theta_{l}^{t}, \phi_{l}^{t}\right)e^{-\jmath\frac{2\pi l \tilde{k}}{\tilde{K}}},
\end{split}
\end{equation}
and $\hat{\textbf{s}}_l^k[\tilde{k}]$ represents the digital precoded signal vector when $\textbf{F}_{RF}^k$ is activated. We assume that the sub-channels corresponding to all sub-carriers of the same rank, i.e., $m=\mathrm{rank}(\textbf{H}[1])=\mathrm{rank}(\textbf{H}[2])=\cdots=\mathrm{rank}(\textbf{H}[\tilde{K}])$. The number of signal vector combinations for $N_{RF}$ data streams per sub-carrier equals to $\left(m\atop N_{RF}\right)$.
Because  the transmissions over all carriers share the same analog precoders \cite{Guo2019}, set $\mathcal{F}_{RF}=\{\textbf{F}_{RF}^1,\textbf{F}_{RF}^2,\cdots,\textbf{F}_{RF}^{K}\}$ can be obtained by solving
\begin{equation}
\begin{split}
(\textbf{P9}):\min_{\{\textbf{F}_{RF}^{k}\},\{\textbf{F}_{BB}^{k}[\tilde{k}]\}}~~&\sum_{\tilde{k}=0}^{\tilde{K}-1}||\textbf{F}_k[\tilde{k}]-\textbf{F}_{RF}^{k}\textbf{F}_{BB}^{k}[\tilde{k}]||_{F}^2\\
\mathrm{subject~to:}~~& \textbf{F}_{RF}^{k}\in \mathcal{U}^{N_t\times N_{RF}},\\&\textbf{F}_{BB}^k[\tilde{k}]\in \mathbb{C}^{N_{RF}\times N_{RF}},\\
&|| \textbf{F}_{RF}^{k}\textbf{F}_{BB}^{k}[\tilde{k}]||_F^2=|| \textbf{F}_{k}[\tilde{k}]||_F^2,
\end{split}
\end{equation}
where $\textbf{F}_k[\tilde{k}]$ is a matrix composed by $N_{RF}$ singular vectors of $\textbf{H}[\tilde{k}]$.
The algorithms for solving (\textbf{P9}) can also be found in rich literature, such as \cite{Ayach2014,Yu2016} and \cite{Sohrabi2016}. Based on $\mathcal{F}_{RF}$ and $\textbf{H}[k]$, we can use the proposed signal shaping method to design  $\mathcal{X}_{N}[k]$ for each sub-carrier.

\begin{figure*}
\setcounter{equation}{58}
\begin{equation}\label{H44}
\textbf{H}=\left[
 \begin{matrix}
  0.2274 - 0.3324\jmath&   0.6728 + 0.4259\jmath&  -0.6300 - 0.9119\jmath&   1.1798 + 0.5234\jmath\\
  -0.5838 + 1.1369\jmath&  -0.0901 - 1.0357\jmath&  -0.1849 + 0.6814\jmath&  -0.4524 - 0.0135\jmath\\
   0.3709 - 0.2147\jmath&   0.6403 + 0.1256\jmath&  -0.9033 - 0.6788\jmath&   1.4362 + 0.3338\jmath\\
  -0.4873 + 1.0504\jmath&  -0.2734 - 0.8536\jmath&   0.0812 + 0.6987\jmath&  -0.6052 - 0.0623\jmath
  \end{matrix}
  \right].
\end{equation}
  \centering
  \includegraphics[width=0.7\textwidth]{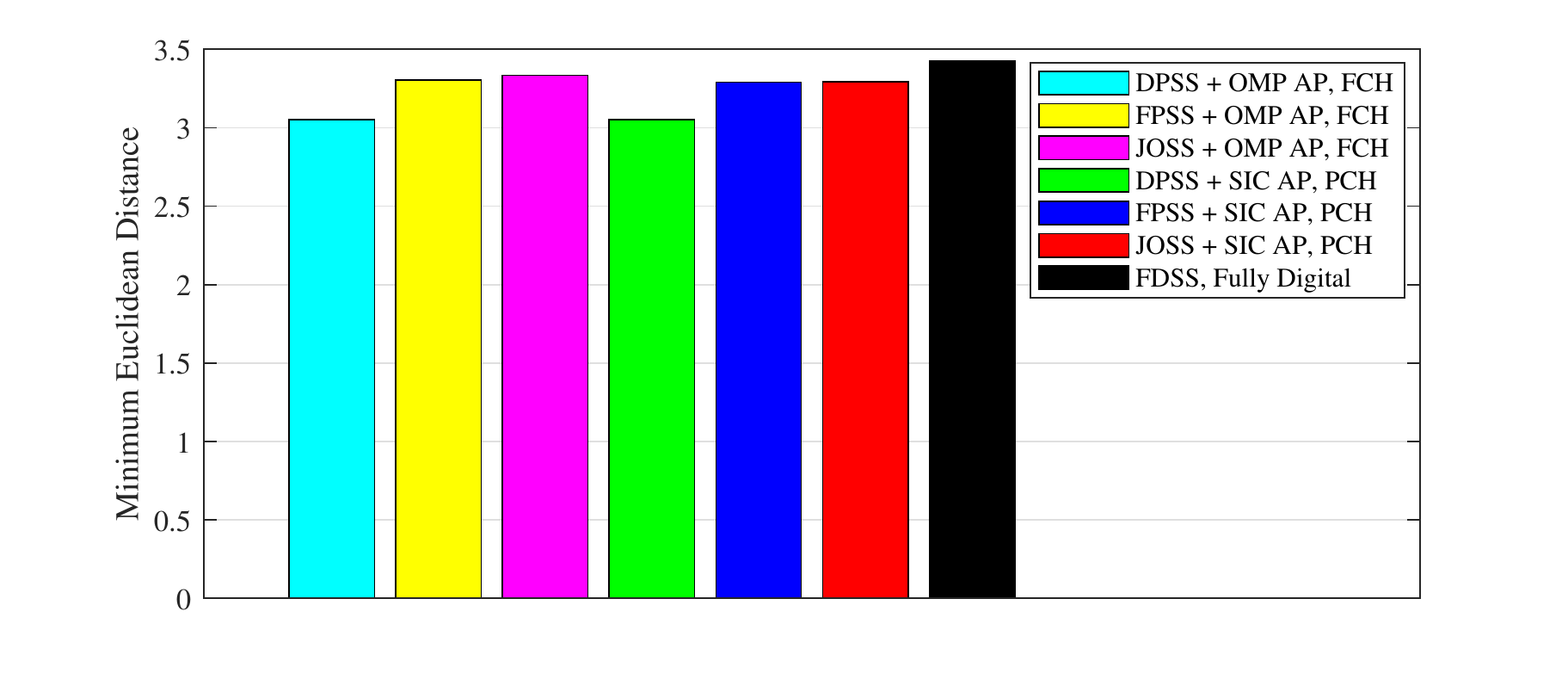}\\
 \caption{Minimum Euclidean distances among the noise-free received vectors when different signal shaping methods are applied in a $(4,4,2,3,3)$ mmWave MIMO system over a constant channel.}
  \label{result0}
\end{figure*}
\subsection{Performance Loss}
To measure the performance loss of the proposed designs, we resort to comparing the proposed signal shaping with FDSS, which is obtained by directly solving (\textbf{RP1}). The FDSS can be obtained as follows. Firstly, we introduce $\textbf{o}_i\in\mathbb{C}^{NN_t}$ as $\textbf{\textbf{o}}_i\triangleq\textbf{g}_i\otimes \textbf{1}_{N_t}$ and  rewrite $\textbf{x}_i$ as
\setcounter{equation}{43}
\begin{equation}\label{xi}
\textbf{x}_i=\hat{\textbf{G}}\hat{\textbf{D}}_\textbf{z}\hat{\textbf{o}}_i,
\end{equation}
where $\hat{\textbf{G}}$ is a matrix  of dimension ${N_r\times N N_t}$ expressed as
\begin{equation}
\hat{\textbf{G}}=[\overbrace{\textbf{H},\textbf{H},\cdots,\textbf{H}}^{N}],
\end{equation}
the matrix $\hat{\textbf{D}}_{\textbf{z}}\in\mathbb{C}^{NN_{t}\times NN_{t}}$ is a diagonal matrix represented by
\begin{equation}
\hat{\textbf{D}}_{\textbf{z}}\triangleq 
\begin{bmatrix}
    ~\textbf{X}_1 & \textbf{0} &\cdots&\textbf{0}\\
    ~\textbf{0} & ~\textbf{X}_2& \cdots& \textbf{0}~ \\
~\vdots& \vdots &  \ddots& \vdots~\\
    ~\textbf{0} &0 &\cdots &\textbf{X}_{N}\\
\end{bmatrix},
\end{equation} 
and
\begin{equation}
\textbf{X}_i=\diag\{\textbf{x}_i\}.
\end{equation}

Based on the similar reformulation illustrated in Sections IV-A and IV-B, we can optimize $\mathcal{X}_{N}=\{\textbf{x}_1,\textbf{x}_2,\cdots,\textbf{x}_{N}\}$ directly by solving
\begin{equation}
\begin{split}
(\textbf{P10}):~\mathrm{Given}: &~\hat{\textbf{Z}}_{ii'}, \forall i\neq i'\in\{1,2,\cdots,N\}\\
\mathrm{Find}:&~\hat{\textbf{z}}\\
\mathrm{Minimize}:&~\hat{\textbf{z}}^T\hat{\textbf{z}}\\
\mathrm{Subject~to}:&~\hat{\textbf{z}}^T\hat{\textbf{Z}}_{ii'}\hat{\textbf{z}}\geq \tau,\forall i\neq i'\in\{1,2,\cdots,N\},
\end{split}
\end{equation}
where $\hat{\textbf{z}}= \diag\{\hat{\textbf{D}_\textbf{z}}\}\in\mathbb{C}^{NN_{t}}$, $\hat{\textbf{Z}}_{i{i'}}=\hat{\textbf{R}}_{\hat{\textbf{G}}}\odot\Delta\hat{\textbf{E}}_{i{i'}}^T\in\mathbb{C}^{NN_t\times NN_{t}}$, $\hat{\textbf{R}}_{\hat{\textbf{G}}}=\hat{\textbf{G}}^H\hat{\textbf{G}}$ and $\Delta\hat{\textbf{O}}_{i{i'}}=(\hat{\textbf{o}}_i-\hat{\textbf{o}}_{i'})(\hat{\textbf{o}}_i-\hat{\textbf{o}}_{i'})^H$. 
It is a QCQP problem with $NN_t$ variables and $\left(N\atop 2\right)$ constraints. The corresponding computational complexity is around $\mathcal{O}(N_{\mathrm{iter}}^{\mathrm{FD}}N^4N_{t}^2)$, where $N_{\mathrm{iter}}^{\text{FD}}$ denotes the average iteration number, by which the algorithm in \cite{Cheng2018} takes to converge to solve (\textbf{P10}). The comparison between FDSS with the hybrid signal shaping methods, including JOSS, FPSS and DPSS, is also included in Table III for clearly viewing. FDSS can be adopted as a benchmark to measure the performance of the proposed signal shaping method for mmWave hybrid MIMO systems.

\subsection{Implementation Challenges}
The proposed signal shaping is designed at each coherent time. After that, the output results can be saved and the signal shaping will be performed according to the saved results at each symbol time. In the high-symbol-rate mmWave communications, the digital part in the signal shaping can be efficiently performed. Therefore, the switching speed of analog precoders is the key factor that determines whether beamspace modulation schemes can be realized in practical mmWave communications. To address this concern, Wang and Zhang have researched the switching speed of analog phase shifters in \cite{Wang2018}. Specifically, there are four types of phase shifters, which are semiconductor, ferroelectric, ferrite,
and micro-electromechanical phase shifters. \cite{Romanofsky2007} showed that the switching slots of semiconductor and ferroelectric phase shifters are in the order of nanosecond. A low-cost  phase shifter design with tens of nanoseconds switching time was reported in \cite{Co2006}. Thanks to these hardware developments, the challenge can be well addressed for high-rate transmissions.

The other implementation challenge is the computational complexity. Even though the FPSS and DPSS have greatly reduced the computational complexity, the complexity still increases with the second power of the number of the analog precoders. To reduce the computational complexity, we can use part of beamspace activation states, where several strong beamspace activation states are selected. By this way, the computational complexity can be reduced. One extreme case is that only the strong beamspace activation state is selected, the signal shaping is reduced to BBSS.

\subsection{Hybrid Receiver-Aware Design}
Considering the signals are typically processed by a hybrid receiver, we can re-express the signal model as
\begin{equation}
\hat{\textbf{y}}=\textbf{W}_{BB}^H\textbf{W}_{RF}^H\textbf{H}\textbf{x}+\textbf{W}_{BB}^H\textbf{W}_{RF}^H\textbf{n}.
\end{equation}
In such a system, the hybrid receiver can be firstly designed by
\begin{equation}
\begin{split}
(\textbf{P11}):\min_{\textbf{W}_{RF},\textbf{W}_{BB}}&~||\textbf{W}_r-\textbf{W}_{RF}\textbf{W}_{BB}||_{F}^2\\
\mathrm{subject~to:}&~\textbf{W}_{RF}\in \mathcal{U}^{ N_r\times N_{RF}^r},~\textbf{W}_{BB}\in \mathbb{C}^{  N_{RF}^{r}\times {m}}.\\
\end{split}
\end{equation}
where $\textbf{W}_r\in\mathbb{C}^{N_{r}\times {N_{RF}^r}}$ is a matrix combined by right-singular vectors; $\textbf{W}_{BB}\in \mathbb{C}^{N_{RF}^{r}\times N_{RF}^{r}}$  represents the digital combiner; $\textbf{W}_{RF}\in \mathcal{U}^{ N_r\times N_{RF}^r}$ stands for the analog combiner and  $N_{RF}^{r}$ is the number of receive RF chains. The problem can be solved by existing algorithms developed in  \cite{Ayach2014,Yu2016,Dai2015,Han2015,Park2017}.  Then, by replacing $\textbf{H}$ with $\textbf{W}_{BB}^H\textbf{W}_{RF}^H\textbf{H}$ and employing the proposed signal shaping methods, we can obtain the hybrid receiver-aware signal shaping.

\subsection{Extension to MSER and MMI Signal Shaping}
With a maximum-likelihood (ML) detector employed at the receiver, the SER of mmWave MIMO systems is upper bounded by \cite{Guo2019b}
\begin{equation}\label{PsX}
\overline{P_{s}}(\mathcal{X}_N)=\frac{1}{2N}\sum_{i=1}^{N}\sum_{i'=1,\atop i'\neq i}^{N}\exp\left(-\frac{\rho}{4}||\textbf{H}(\textbf{x}_i-\textbf{x}_{i'})||_2^2\right),
\end{equation}
where $\rho$ represents the SNR.
Based on the  re-formulation in Sections IV-A, we have 
\begin{equation}
||\textbf{H}(\textbf{x}_i-\textbf{x}_{i'})||_2^2=\textbf{z}^T\textbf{Z}_{i{i'}}\textbf{z},
\end{equation}
and the upper bound can be re-expressed as
\begin{equation}\label{PsMetric}
\overline{P_{s}}(\textbf{z})=\frac{1}{2N}\sum_{i=1}^{N}\sum_{i'=1,\atop i'\neq i}^{N}\exp\left(-\frac{\rho}{4}\textbf{z}^T\textbf{Z}_{i{i'}}\textbf{z}\right).
\end{equation}
Thus, the MSER signal shaping can be formulated as
\begin{equation}
\begin{split}
(\textbf{SER-OP}):~~\mathrm{Given}: &~\textbf{Z}_{ii'}, \forall i\neq i'\in\{1,2,\cdots,N\}, \rho\\
\mathrm{Find}:&~\textbf{z}\\
\mathrm{Minimize}:&~\overline{P_{s}}(\textbf{z})\\
\mathrm{Subject~to}:&~\textbf{z}^T\textbf{z}\leq N.
\end{split}
\end{equation}
 Given $\mathcal{X}$ as inputs, the mutual information of mmWave MIMO systems can be written as \cite{Guo2014}
\begin{equation}
\begin{split}
&\mathcal{I}(\textbf{x};\textbf{y}|\textbf{H})=\log_2 N-\cdots\\
&\frac{1}{N}\sum_{i=1}^{N}\mathbb{E}_{\textbf{n}}\left\{\log_2\sum_{i'=1}^N\exp\left[-\rho({||\textbf{H}(\textbf{x}_i-\textbf{x}_{i'}+\textbf{n}^2)||_2^2-||\textbf{n}||_2^2})\right]\right\},
\end{split}
\end{equation}
which is lower bounded by \cite{Zeng2012}
\begin{equation}\label{MIX}
\begin{split}
\mathcal{I}_{LB}(\textbf{x};\textbf{y}|\textbf{H})=&\log_2N+N_r(1-\log_2e)\\
&-\frac{1}{N}\sum_{i=1}^{N}\log_2\sum_{i'=1}^{N}\exp\left(-\frac{\rho||\textbf{H}(\textbf{x}_i-\textbf{x}_{i'}||_2^2}{2}\right).
\end{split}
\end{equation}
With the reformulations
$||\textbf{H}(\textbf{x}_i-\textbf{x}_{i'})||_2^2=\textbf{z}^T\textbf{Z}_{i{i'}}\textbf{z}$ when $i\neq i'$, the mutual information lower bound can be expressed as
\begin{equation}\label{MIMetric}
\begin{split}
\mathcal{I}_{LB}(\mathbf{q})=&\log_2N+N_r(1-\log_2e)\\
&-\frac{1}{N}\sum_{i=1}^{N}\log_2\left[1+\sum_{{i'=1}\atop{i'\neq i}}^{N}\exp\left(-\frac{\rho\textbf{z}^T\textbf{Z}_{i{i'}}\textbf{z}}{2}\right)\right].
\end{split}
\end{equation}
Thus, the MMI signal shaping can be formulated as
\begin{equation}
\begin{split}
(\textbf{MI-OP}):~~\mathrm{Given}: &~\textbf{Z}_{ii'}, \forall i\neq i'\in\{1,2,\cdots,N\}, \rho\\
\mathrm{Find}:&~\textbf{z}\\
\mathrm{Maximize}:&~\mathcal{I}_{LB}(\mathbf{z})\\
\mathrm{Subject~to}:&~\textbf{z}^T\textbf{z}\leq N
\end{split}
\end{equation}
By replacing (\textbf{P5}) with (\textbf{SEP-OP}) and (\textbf{MI-OP}), the JOSS approach can be directly extended to the designs based on the MSER and MMI criteria, respectively.  (\textbf{SEP-OP}) and (\textbf{MI-OP}) can also be solved by the existing algorithms, e.g., the algorithm designed in \cite{Wang2015}. The computational complexity are of the same order as that for solving (\textbf{P5}), because the objective functions of (\textbf{SEP-OP}) and (\textbf{MI-OP}) are the functions of $\{\textbf{z}\textbf{Z}_{ii'}\textbf{z}\}$, whose computation dominates the computational complexity. In a similar way,  FPSS, DPSS, and FDSS can be extended.
It should be remarked that the MSER and MMI criteria are equivalent to  the MMED criterion in the high SNR regime. The reason is that the upper bound on SER given (\ref{PsMetric}) and the lower bound on MI given in (\ref{MIMetric}) are dominated by the minimum Euclidean distance term  with the help of the exponential operator. Moreover, it should be noted that  MMED design is invariant to the instantaneous SNR,  while the MSER and MMI designs have to be updated as SNR varies.

\section{Simulation and Analysis}
In this section, we will present simulation results to show  the superiority of the proposed signal shaping methods aided NUBM and to validate its effectiveness in mmWave broadband systems, in the systems with channel estimation errors and that with hardware impairments.  For clear denotation, we use the parameters $(N_t,N_r,N_{RF},n,L)$ to characterize an mmWave FCH/PCH MIMO system.

\subsection{Performance over a Constant mmWave MIMO Channel}
Considering the time consumption for the CSI acquisition and the computation latency in the design procedure, the proposed signal shaping methods are more suitable for mmWave MIMO communications that experience slowly varying channels. The potential applications can be wireless backhaul communications,  wireless big data communications for data centers, and in-car/in-device high-rate data communications. Thus, we first investigate the application over a constant mmWave MIMO channel.

 In a $(4,4,2,3,3)$ mmWave MIMO system over a constant channel given in (\ref{H44}), we demonstrate the minimum Euclidean distance of the proposed signal shaping methods for mmWave hybrid MIMO systems with different structures and analog precoding (AP) methods in Fig. \ref{result0}. In particular, we compare DPSS, FPSS, and JOSS in FCH MIMO with OMP AP \cite{Ayach2014} and that in PCH MIMO with SIC AP \cite{Dai2015}. FDSS with a fully-digital structure is also included as a benchmark. Simulation results show that the proposed hybrid JOSS and FPSS methods can approximately approach the FDSS with a fully-digital structure, i.e., obtaining almost the same minimum Euclidean distance. This indicates that solving (\textbf{P2}) and (\textbf{P3}) to obtain the solution to (\textbf{P1}) is an effective way. Observing our methods of solving (\textbf{P3}),  it is found that JOSS and FPSS outperform DPSS. Specifically, the achieved minimum Euclidean distances by JOSS and FPSS are around $10\%$ higher than that achieved by DPSS. 
   It is also shown that PCH with the proposed signal shaping methods can achieve similar performance with the FCH structure. This is because the number of transmit antennas is small. When the number of transmit antennas becomes larger, the difference between the PCH and the FCH structures will also become larger. 
 
\subsection{Performance in  mmWave Massive FCH MIMO Systems}
\begin{figure}[t]
  \centering
  \includegraphics[width=0.5\textwidth]{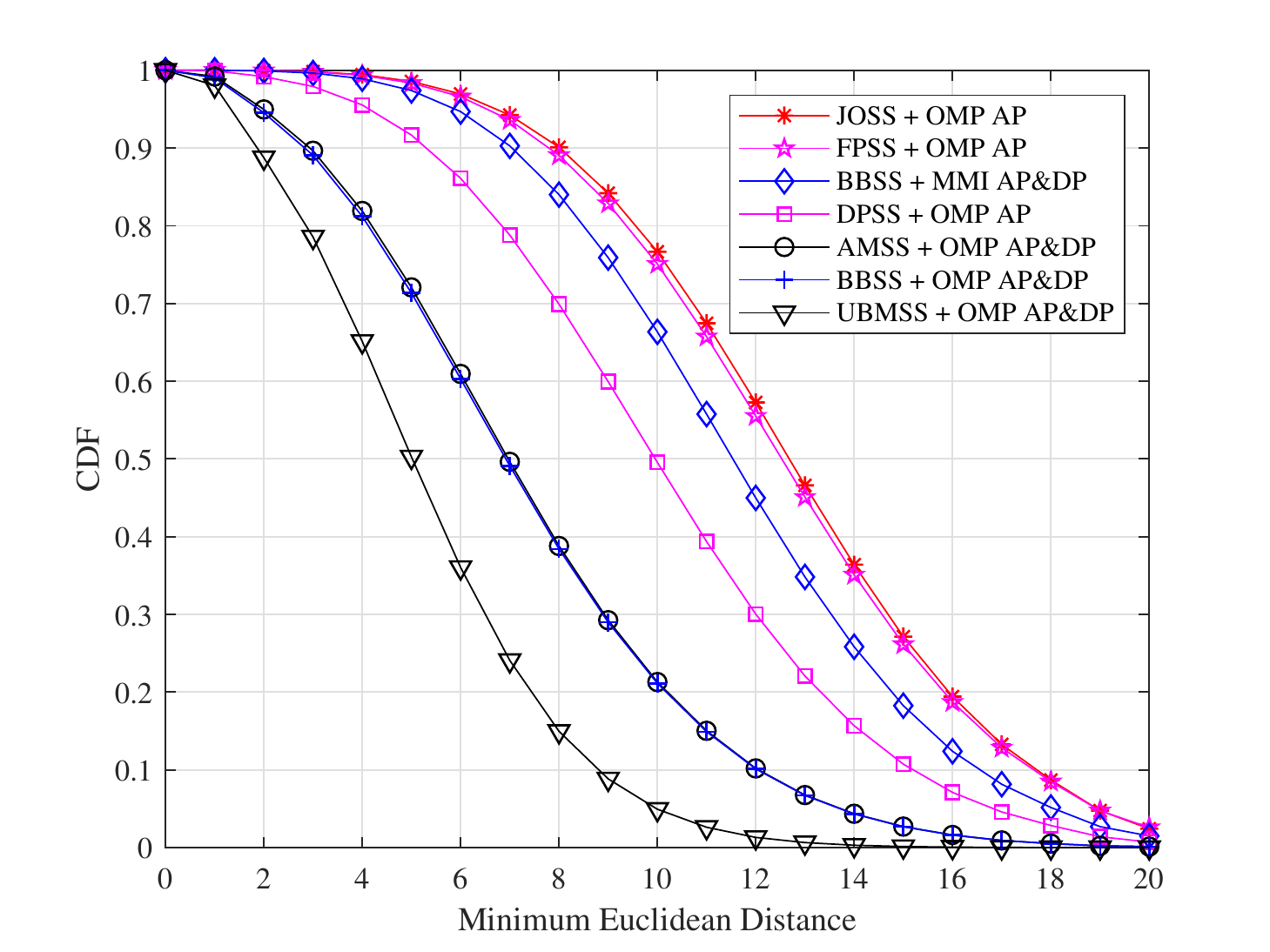}\\
 \caption{CDF of the minimum Euclidean distances among the noise-free received vectors when different signal shaping methods are applied in a $(64,4,2,3,3)$ mmWave FCH MIMO system.}
  \label{result1a}
\end{figure}

\begin{figure}[t]
  \centering
  \includegraphics[width=0.5\textwidth]{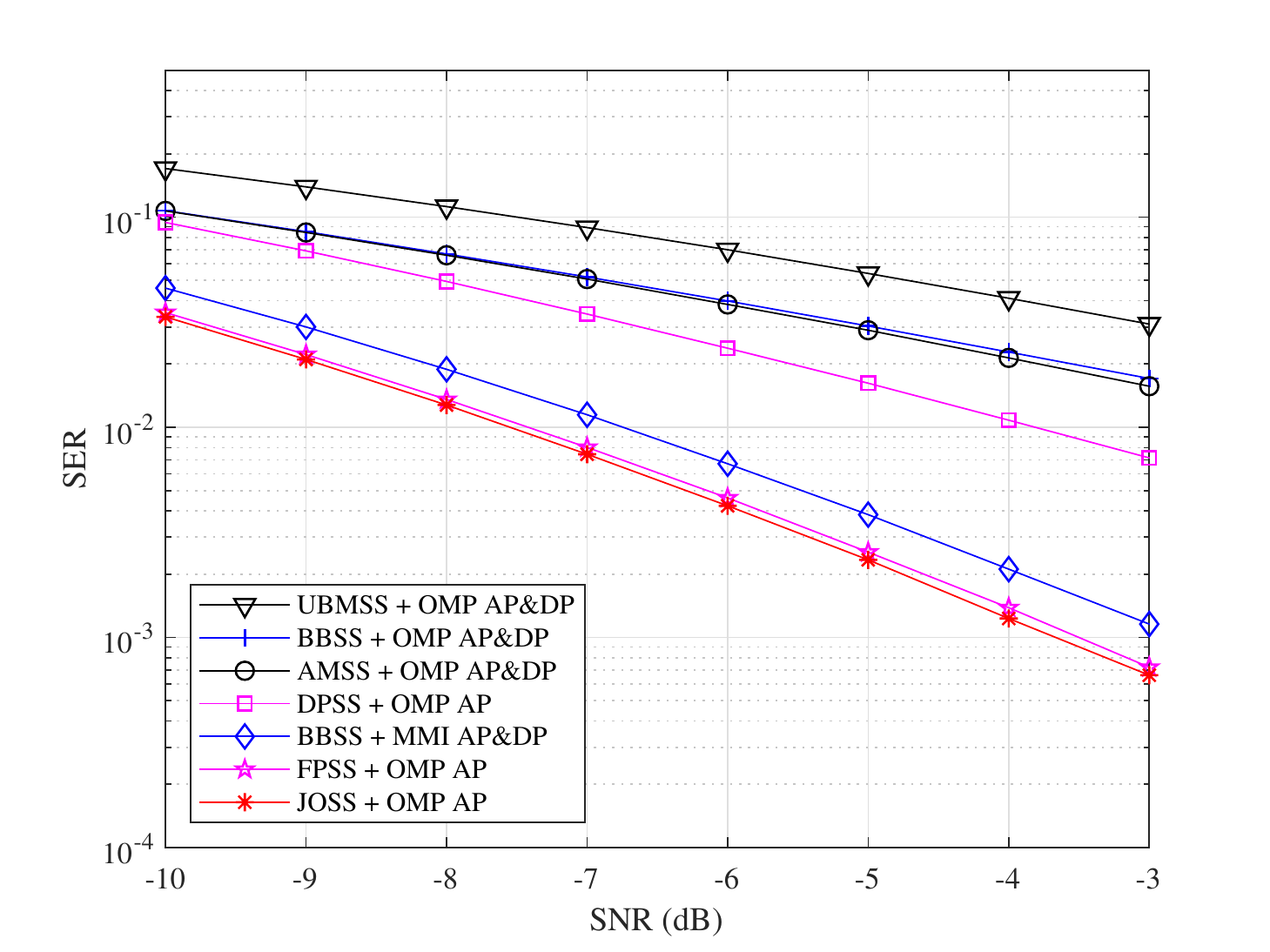}\\
 \caption{SER of different signal shaping methods in a $(64,4,2,3,3)$ mmWave FCH MIMO system.}
  \label{result1b}
\end{figure}

In a $(64,4,2,3,3)$ mmWave FCH MIMO system, we evaluate the cumulative distribution function (CDF) of the minimum Euclidean distances among all noise-free received vectors, SER and the computational complexity quantified by the number of floating operations of various schemes as illustrated in Figs. \ref{result1a}-\ref{result1b}, and Table IV. In detail, seven signal shaping schemes are compared in the system setup. ``UBMSS + OMP AP\&DP" represents the UBMSS scheme with analog precoders and digital precoders (AP\&DP) generated by OMP algorithm \cite{Ayach2014}. ``BBSS + OMP AP\&DP" and ``AMSS + OMP AP\&DP" stand for the BBSS scheme with OMP AP\&DP and adaptive modulation-based signal shaping (AMSS)\cite{Gao2019} with OMP AP\&DP, respectively. ``BBSS+ MMI AP\&DP" represents the BBSS scheme with MMI AP\&DP \cite{Rajashekar2016}, which is designed by assuming finite alphabet inputs. Observing the comparison results in Figs. \ref{result1a} and \ref{result1b}, we find that JOSS exhibits the best performance and slightly outperforms FPSS. Both of them outperform existing transmission solutions for mmWave MIMO communications.
The proposed DPSS enjoys lower computational complexity compared to FPSS and JOSS, but the reduction of complexity reduces the system performance greatly.  The performance loss of DPSS compared to FPSS mainly comes from the limited symbol vector set refinement capability, because only the diagonal elements in the precoding matrix can be adjusted. Observing the comparison in computational complexity quantified by the number of operations in Table IV,  as expected, the proposed signal shaping involves additional symbol vector optimization for performance improvement and incurs much computational complexity. However, the complexity is acceptable for mmWave MIMO communications, in which the transmit signals propagate over slow fading channels.

\begin{table}[t] 
\centering
\caption{Computational Complexity of Different Signal Shaping Methods by the Number of Floating-Point Operations in a $(64,4,2,3,3)$ mmWave FCH MIMO System.} 
\begin{tabular}{ | c || c |}
\hline
\textbf{Signal Shaping Methods} &  \textbf{Computational Complexity} \\ 
\hline
JOSS + OMP AP &$4.89\times 10^7$\\
\hline
FPSS + OMP AP & $7.45\times 10^6$\\
\hline
BBSS + MMI AP\&DP & $4.82\times 10^6$\\
\hline
DPSS + OMP AP & $3.21\times 10^6$\\
\hline
AMSS + OMP AP\&DP & $5.03\times 10^4$\\
\hline
BBSS + OMP AP\&DP & $2.31\times 10^4$\\
\hline
UBMSS + OMP AP\&DP & $4.92\times 10^4$\\
\hline
\end{tabular}
\end{table}
 
\subsection{Performance in mmWave Massive  PCH MIMO Systems}
\begin{figure}[t]
  \centering
  \includegraphics[width=0.5\textwidth]{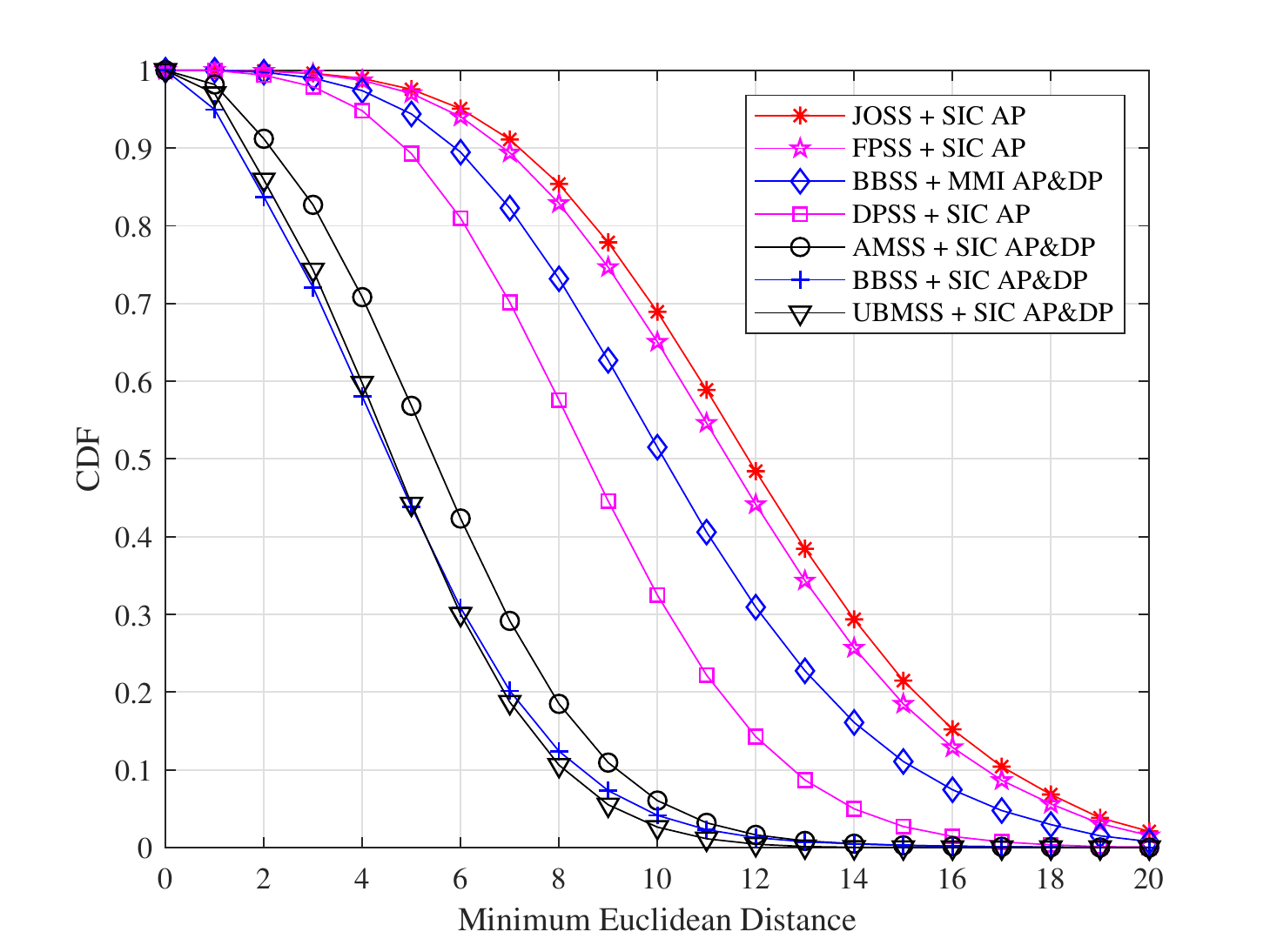}\\
 \caption{CDF of the minimum Euclidean distances among the noise-free received vectors when different signal shaping methods are applied in a $(64,4,2,3,3)$ mmWave PCH MIMO system.}
  \label{result3a}
\end{figure}

\begin{figure}[t]
  \centering
  \includegraphics[width=0.5\textwidth]{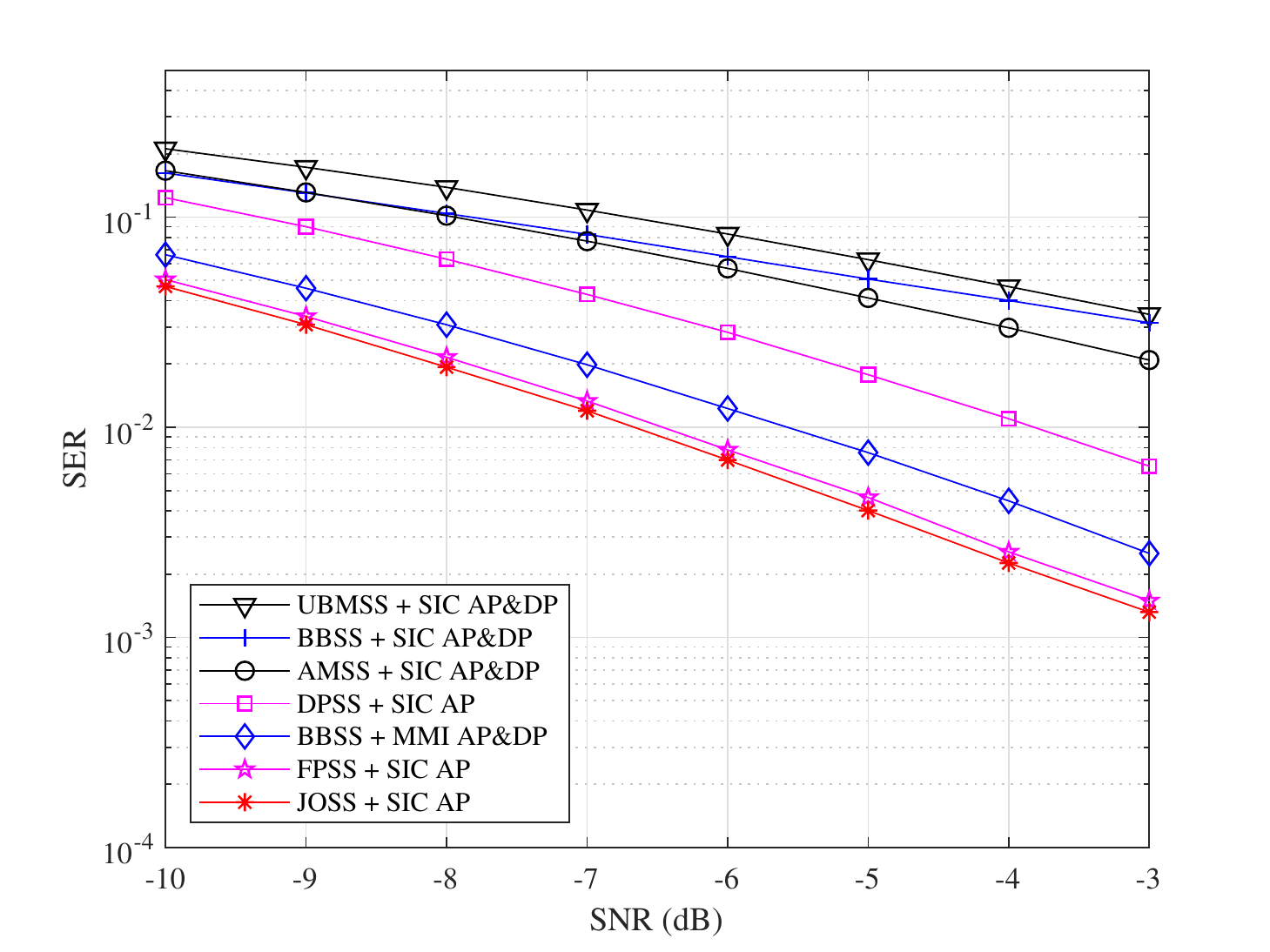}\\
 \caption{SER of different signal shaping methods in a $(64,4,2,3,3)$ mmWave PCH MIMO system.}
  \label{result3b}
\end{figure}
Besides the FCH MIMO system, we also make the comparisons in a PCH MIMO system as illustrated in Figs. \ref{result3a}-\ref{result3b}, and Table V. In detail, seven signal shaping schemes are compared in the system setup. ``UBMSS +SIC AP\&DP" represents the UBMSS scheme with AP\&DP generated by SIC algorithm \cite{Dai2015}. ``BBSS + SIC AP\&DP" and ``AMSS + SIC AP\&DP" stand for the BBSS scheme with SIC AP\&DP and AMSS \cite{Gao2019} with SIC AP\&DP, respectively.  From comparison results, we can draw similar conclusion that the proposed JOSS and FPSS outperforms existing mmWave MIMO transmission solutions. DPSS can only outperform existing signal shaping methods, which are obtained by assuming complex Gaussian inputs. Besides, by jointly observing the results in Figs. \ref{result1b} and \ref{result3b}, one can find that FCH MIMO with the proposed signal shaping greatly outperforms PCH MIMO with the proposed signal shaping. The better performance of FCH MIMO results from the higher beamforming gain of FCH MIMO.

\begin{table}[t] 
\centering
\caption{Computational Complexity of Different Signal Shaping Methods by the Number of Floating-Point Operations in a $(64,4,2,3,3)$ mmWave PCH MIMO System.} 
\begin{tabular}{ | c || c |}
\hline
\textbf{Signal Shaping Methods} &  \textbf{Computational Complexity} \\ 
\hline
JOSS + SIC AP &$1.98\times 10^7$\\
\hline
FPSS + SIC AP & $3.01\times 10^6$\\
\hline
BBSS + MMI AP\&DP & $2.53\times 10^6$\\
\hline
DPSS + SIC AP &  $1.61\times 10^6$\\
\hline
AMSS + SIC AP\&DP & $3.51\times 10^5$\\
\hline
BBSS + SIC AP\&DP & $9.78\times 10^4$\\
\hline
UBMSS + SIC AP\&DP & $3.43\times 10^5$\\
\hline
\end{tabular}
\end{table}

\subsection{Performance in mmWave MIMO-OFDM Systems}
\begin{figure}[t]
  \centering
  \includegraphics[width=0.5\textwidth]{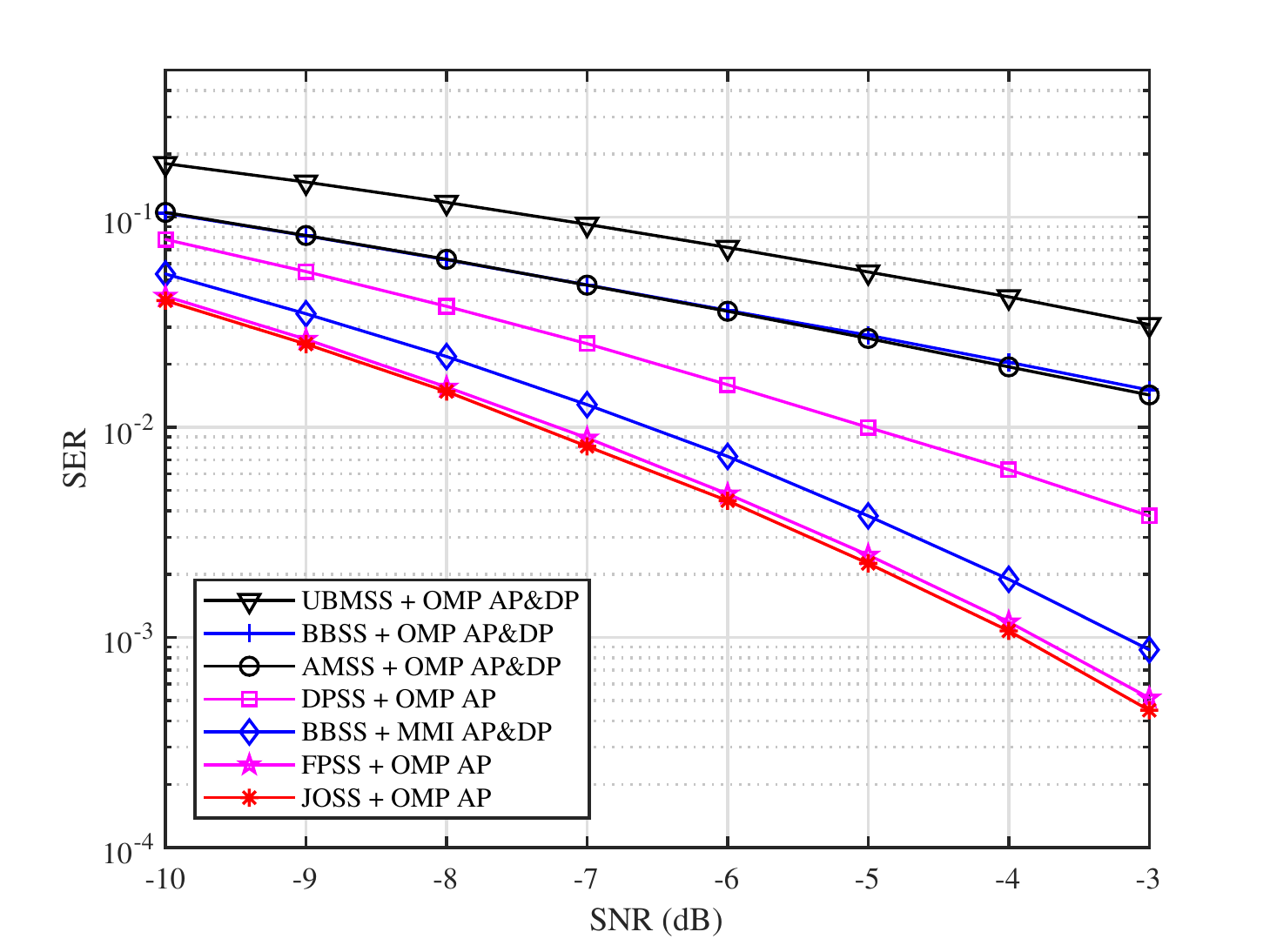}\\
 \caption{SER of different signal shaping methods in a $(16,9,2,3,3)$ mmWave FCH MIMO-OFDM system with 128 sub-carriers.}
  \label{result5}
\end{figure}

For broadband mmWave MIMO communications, we simulate the SER of a $(16,9,2,3,3)$ mmWave FCH MIMO-OFDM system with 128 sub-carriers. Similarly, we compare NUBM using FPSS, DPSS and JOSS with UBMSS and BBSS. The results are illustrated in Fig. \ref{result5}, from which it is validated the proposed signal shaping can be easily extended to broadband communication to bring considerable gain. Specifically, NUBM with FPSS and JOSS outperforms BBSS with MMI AP\&DP by around 0.8 dB. They outperform  AMSS and UBMSS with OMP AP by 5 dB and 10 dB, respectively.


\subsection{Performance in the Presence of Channel Estimation Errors}
\begin{figure}[t]
  \centering
  \includegraphics[width=0.5\textwidth]{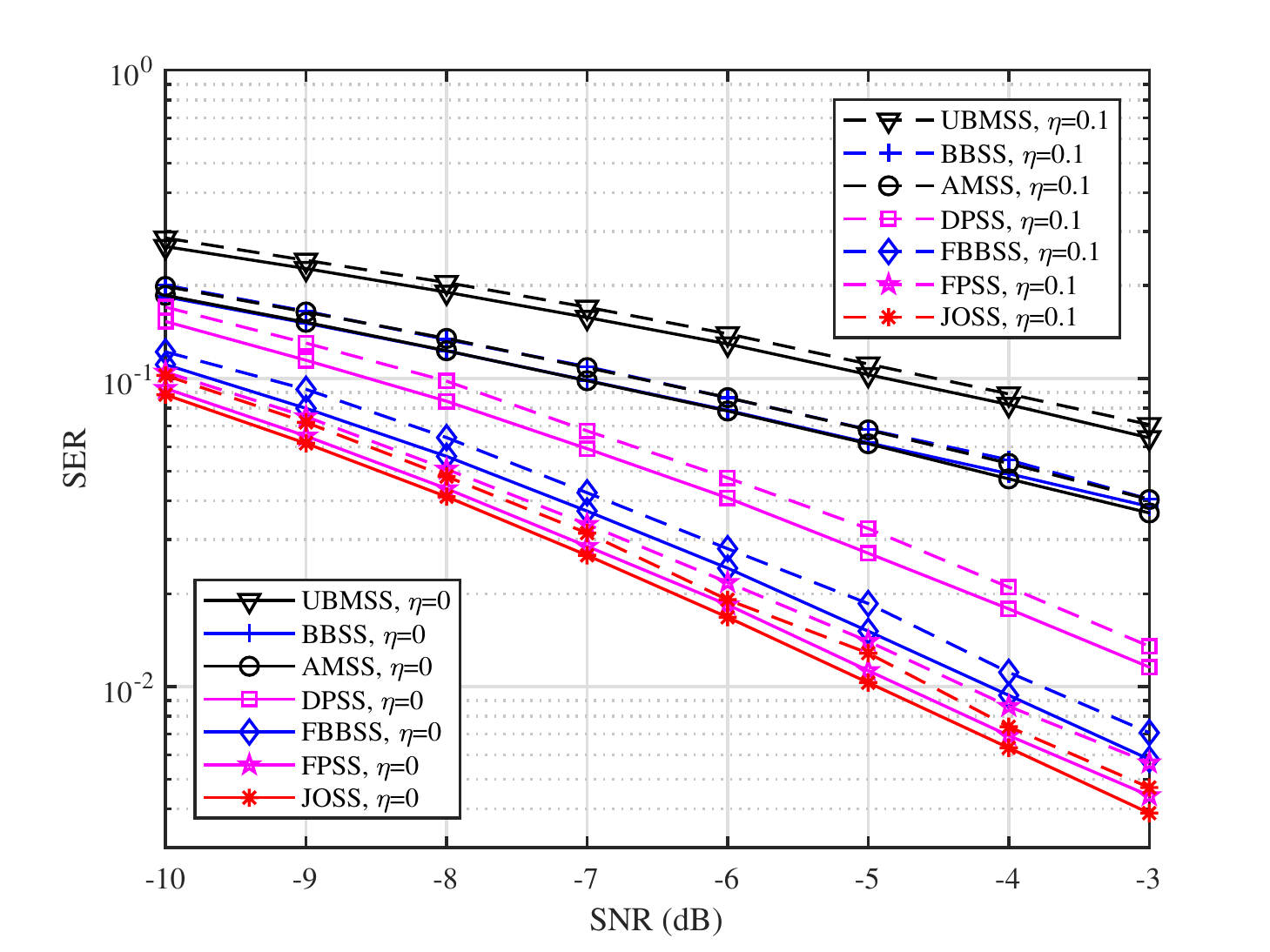}\\
 \caption{SER of different signal shaping methods in a $(36,4,2,3,3)$ mmWave FCH MIMO system with perfect CSI ($\eta=0$) and imperfect CSI ($\eta=0.1$).}
  \label{result4}
\end{figure}

All of the designs are based on the perfect CSI at the transceivers. To show the robustness in the presence of channel estimation errors, a simplified channel error model $\textbf{H}_{\rm{im}}=\textbf{H}+\textbf{H}_e$ \cite{Guo2016,Guo2017a} is adopted, where $\textbf{H}_e$ denotes the matrix of channel estimation errors with each entry obeying a complex Gaussian distribution with zero mean and variance $\sigma_e^2$; $\sigma_e^2$ is propositional to the variance of the noise, i.e., $\sigma_e^2=\eta\sigma_n^2$. It is noteworthy that the channel estimation error model is just an example demonstrating the worst case. In the simulation, we set $\eta=0.1$ and compare the simulation results with that using perfect CSI, as illustrated in Fig. \ref{result4}. Simulation results demonstrate that all schemes experience similar performance losses in the presence of imperfect CSI and the proposed JOSS and FPSS maintain the superiority over other schemes.  In other words, all schemes have similar robustness to channel estimation errors, and in the presence of a similar level of channel estimation errors, the proposed signal shaping methods can achieve the best performance in comparison with existing transmission solutions.

\subsection{Performance in the Presence of Hardware Impairments}
\begin{figure}[t]
  \centering
  \includegraphics[width=0.5\textwidth]{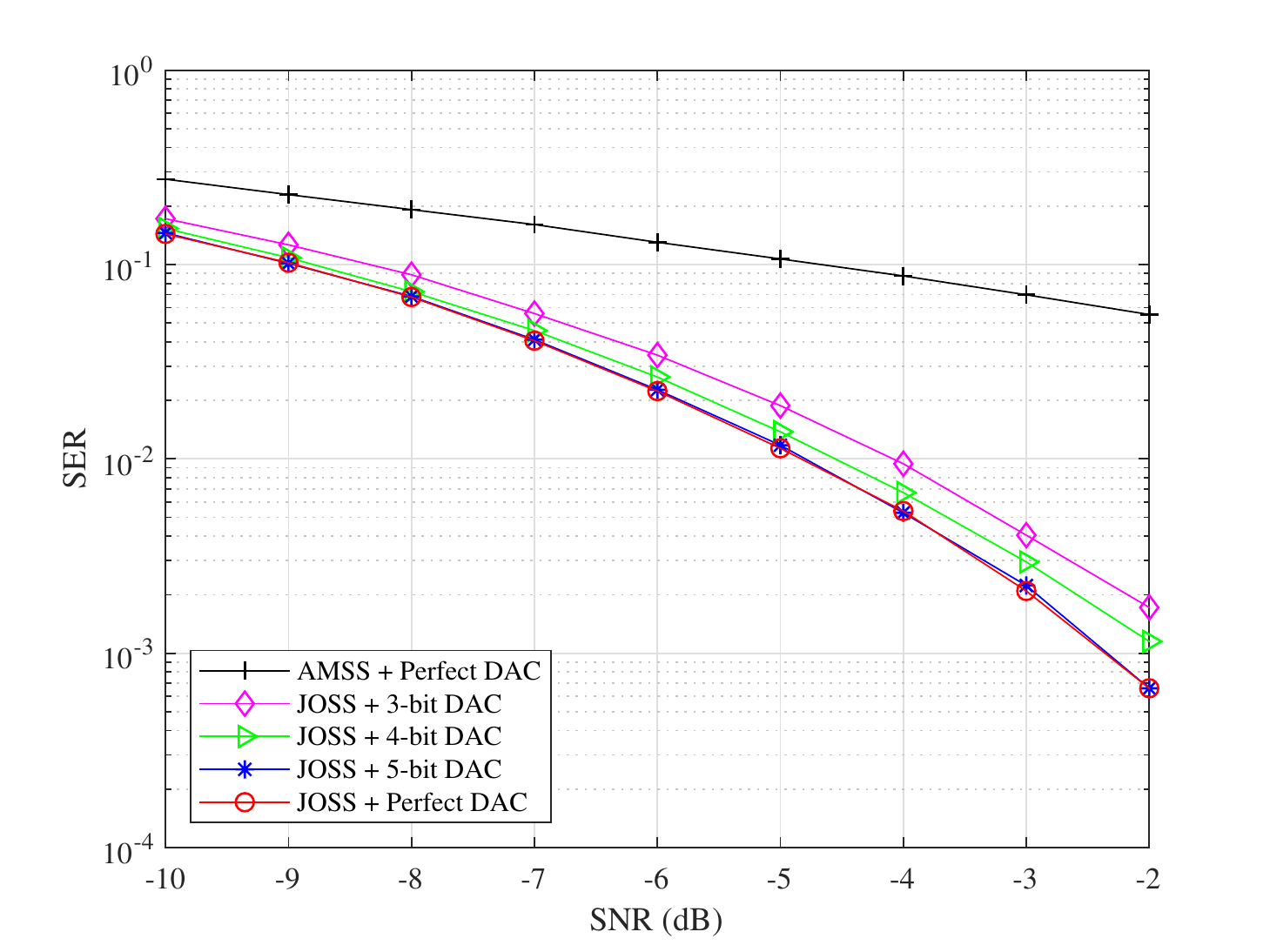}\\
 \caption{SER of the proposed JOSS method in $(16,4,2,3,3)$ mmWave FCH MIMO systems using perfect DAC and imperfect DAC.}
  \label{result9}
\end{figure}

Since  mmWave communication systems could be equipped with imperfect hardware, such as the phase/amplitude inconsistent circuits, low-resolution digital-to-analog converters (DAC) and analog-to-digital converters (ADC). To show the robustness of the proposed designs, we simulate the performance of the proposed JOSS method in mmWave MIMO systems using low-resolution DAC at the transmitter as illustrated in Fig. \ref{result9}. As depicted in Fig. \ref{result9}, the proposed JOSS method maintain good performance when the 5-bit DAC is adopted. When the $4$-bit and $3$-bit DAC are adopted, there are $0.4$ dB and $0.8$ dB performance losses, respectively. Despite that, it can be observed that the achieved performance gain is substantial compared to AMSS.

\section{Conclusion}
For mmWave MIMO communications with CSI at the transmitter, we investigated the signal shaping methods according to the MMED criterion. Different from existing BBSS schemes that only activate fixed beamspace per coherent time and UBMSS schemes that equiprobably activate each beamspace, our proposed methods activate different beamspace with different probabilities and with different symbol vector sets. In other words, our designs are more generalized, which also results in better performance. 

Specifically, in mmWave hybrid MIMO communication systems, we split the transmit vector shaping methods into an analog precoder design problem and a symbol vector set optimization problem. Then, based on existing work on analog precoder optimization, we dedicated our effort to the symbol vector set optimization. We proposed three signal shaping methods: JOSS, FPSS and DPSS. Among them, JOSS optimizes the symbol vector sets for each optimized analog precoder directly, including the set size optimization and set entry optimization. The searching space of JOSS is the largest and thus JOSS is of the highest computational complexity. To reduce the complexity, we adopted full or diagonal precoders to refine predefined symbol vector sets, i.e., FPSS and DPSS. By reducing the optimization search space, the computational complexity is reduced accordingly. Finally, we discussed the proposed signal shaping methods in the applications in OFDM-based mmWave MIMO communications and in mmWave MIMO communications with hybrid transceivers. 

Simulation results revealed that the proposed JOSS and FPSS outperform existing BBSS and UBMSS methods; FPSS exhibits similar performance compared to JOSS but with much lower complexity; DPSS also reduces a lot of complexity but at the cost of significant performance loss. Moreover, simulations also validated that the proposed signal shaping methods can be extended to mmWave MIMO-OFDM systems, mmWave MIMO systems with hybrid transceivers, mmWave MIMO systems with imperfect CSI and hardware impairment. The superiority of the proposed signal shaping maintains in these systems. In summary, the proposed signal shaping methods achieve better performance than BBSS and UBMSS, and can be a promising candidate in the future mmWave MIMO communications.
\bibliographystyle{IEEEtran}
\bibliography{IEEEabrv,hybrid}

\end{document}